# Beyond domain alignment: Revealing the effect of intrinsic magnetic order on electrochemical water splitting

Emma van der Minne,[1,*] Lucas Korol,[2] Lidewij M.A. Krakers,[1] Michael Verhage,[3] Carlos M. M. Rosário,[4] Thijs J. Roskamp,[1] Raymond J. Spiteri,[2] Chiara Biz,[5] Mauro Fianchini,[5] Guus Rijnders,[1] Kees Flipse,[3] Jose Gracia,[5] Guido Mul,[6] Hans Hilgenkamp,[1] Robert J. Green,[2,7] Gertjan Koster,[1] Christoph Baeumer[1,8,*]

**AFFILIATIONS**

[1] MESA+ Institute for Nanotechnology, Faculty of Science and Technology, University of Twente, Enschede 7500AE, Netherlands
[2] Department of Physics & Engineering Physics, University of Saskatchewan, Saskatoon, Canada S7N 5E
[3] Molecular Materials and Nanosystems (M2N) - Department of Applied Physics - Eindhoven University of Technology, Eindhoven, Netherlands
[4] Laboratory for Nanostructured Solar Cells- International Iberian Nanotechnology Laboratory- Braga, Portugal
[5] MagnetoCat SL, 03012 Alicante, Spain
[6] Photocatalytic Synthesis (PCS) group, University of Twente, 7522 NB Enschede, Netherlands
[7] Stewart Blusson Quantum Matter Institute, University of British Columbia, Vancouver, Canada V6T 1Z1
[8] Peter Gruenberg Institute and JARA-FIT, Forschungszentrum Juelich GmbH, Juelich 52425, Germany
* corresponding authors: e.vanderminne@utwente.nl; c.baeumer@utwente.nl

**ABSTRACT**

To reach a long term viable green hydrogen economy, rational design of active oxygen evolution reaction (OER) catalysts is critical. An important hurdle in this reaction originates from the fact that the reactants are singlet molecules, whereas the oxygen molecule has a triplet ground state with parallel spin alignment, implying that magnetic order in the catalyst is essential. Accordingly, multiple experimentalists reported a positive effect of *external* magnetic fields on OER activity of ferromagnetic catalysts. However, it remains a challenge to investigate the influence of the *intrinsic* magnetic order on catalytic activity. Here, we tuned the intrinsic magnetic order of epitaxial $La_{0.67}Sr_{0.33}MnO_3$ thin film model catalysts from ferro- to paramagnetic by changing the temperature in-situ during water electrolysis. Using this strategy, we show that ferromagnetic ordering below the Curie temperature enhances OER activity. Moreover, we show a slight current density enhancement upon application of an external magnetic field and find that the dependence of magnetic field direction correlates with the magnetic anisotropy in the catalyst film. Our work thus suggests that both the intrinsic magnetic order in $La_{0.67}Sr_{0.33}MnO_3$ films and magnetic domain alignment increase their catalytic activity. We observe no long-range magnetic order at the catalytic surface, implying that the OER enhancement is connected to the magnetic order of the bulk catalyst. Combining the effects found with existing literature, we propose a unifying picture for the spin-polarized enhancement in magnetic oxide catalysts.

## INTRODUCTION

To establish a sustainable energy infrastructure, efficient energy storage systems are of the highest interest to overcome the intermittent nature of renewable energy sources. Green hydrogen is one of the most promising fuels for energy storage and an ideal feedstock for coupling renewable energy with other sectors like the chemical and the steel industries. [1,2] However, production of green hydrogen via water electrolysis suffers from sluggish kinetics in the oxygen evolution reaction (OER). [3–5] This might be connected to the fact that the reactants, $OH^-$ or $H_2O$, are diamagnetic, but the final product, triplet $O_2$, is paramagnetic.[6,7]

Accordingly, recent theoretical investigations suggest that the spin polarized orbital configurations in ferromagnetic bonds increase OER efficiency by promoting the generation of triplet oxygen by quantum spin-exchange interactions (QSEI) and intrinsic spin filtering through exchange splitting of the energy levels in the conduction band of a magnetic material. [6–10] From the orbital physics of correlated electrons, it is hypothesized that maximum OER activity can be obtained at catalytic surfaces with dominant metallic ferromagnetic behavior. [9,10]

Following this idea, multiple experimental studies reported a positive effect of external magnetic fields on OER activity of ferromagnetic catalysts, suggesting several possible explanations. [6,11–20] An increase in spin polarization at the active sites was suspected to reduce the charge transfer resistance at the electrode-electrolyte interface and increase spin selective adsorption. [6,16,20] The removal of domain walls may reduce domain wall scattering and thus lower the magnetoresistance and increase the amount of ferromagnetically coupled reactive sites. [14,15] An increase in the ferromagnetic exchange field between antiferromagnetic or paramagnetic catalyst surfaces and ferromagnetic subsurface layers can increase the extent of spin order at the reaction sites. [17,18] Lastly, the magnetoresistance effect may increase the activity by decreasing the electronic resistance in the catalyst. [11] From the lack of significant changes of OER enhancement on purely paramagnetic catalysts it was hypothesized that the effects of Lorentz and Kelvin forces can be excluded as dominant factors for magnetic field enhanced OER activity. [6,12,21]

Although it has been shown that *external* field application on open-shell catalysts can enhance OER activity, it remains a challenge to investigate the influence of the *intrinsic* magnetic order in catalysts. Initial attempts to study the effects of intrinsic magnetic order have shown that a higher saturation magnetization [22,23], the occurrence of spin channels [24], a higher spin magnetic moment on the active sites [25–28], and the introduction of ferromagnetism [26] can enhance either the OER activity directly, or its response to magnetic field exposure. However, to date these changes in magnetic order were accompanied by a change in either the composition, the crystal structure, or the crystal symmetry of the catalyst such that it remains a challenge to pinpoint the observed effects to the intrinsic magnetic order.

In this article, we introduce a strategy to vary the magnetic order without applied magnetic field during catalysis and without changing the crystal structure. We employed epitaxial $La_{0.67}Sr_{0.33}MnO_3$ thin films grown by pulsed laser deposition (PLD) [29] as model catalysts with a Curie temperature ($T_c$) slightly above and below room temperature. This enabled us to change the magnetic order of the catalyst from ferromagnetic to paramagnetic in-situ during water electrolysis by changing the temperature. At $T_c$, a change of catalytic activity is expected connected to the disturbance of the inter-atomic exchange interactions due to thermal fluctuations, also known as the Hedvall effect, as observed for a range of other catalytic processes.[30] Generally, one expects a decrease in current density with decreasing temperature because of smaller thermal energy. By comparing the para- and ferromagnetic films we show that for ferromagnetic films, the current densities below the Curie temperature were higher than expected if only temperature dependent effects are considered. This indicates that the presence of ferromagnetic ordering below the Curie temperature, i.e. the intrinsic magnetic order, indeed enhances OER activity.

The importance of ferromagnetic order is further demonstrated by an enhancement of OER activity for the same ferromagnetic film upon alignment of magnetic domains with an external magnetic field. We show a correlation between the magnetic anisotropy in our catalyst and the external magnetic field enhancement. Our work thus suggests that both the intrinsic magnetic order in $La_{0.67}Sr_{0.33}MnO_3$ films and the magnetic domain alignment upon external field exposure increase the catalytic activity. Moreover, the long-range magnetic order at the catalytic surface is strongly suppressed, implying that the OER enhancement is related with the magnetic ordering of the catalyst bulk, rather than lateral long-range order in the surface layer. Combining our observations with existing literature, we propose a unifying picture for the spin-polarized enhancement in magnetic oxide catalysts.

**RESULTS AND DISCUSSION**

To study the effects of intrinsic magnetic order, one needs a model system in which physical material properties can be changed without changing composition or crystal structure. Epitaxial thin films in which highly controlled synthesis helps finetune physical properties and electrochemical functionality while maintaining composition and structure have emerged as an ideal platform to identify structure-function-relationships at the atomic level. [31,32] Here, we selected $La_{0.67}Sr_{0.33}MnO_3$ which is a half metal at room temperature associated with a ferromagnetic double exchange mechanism from (intra- and inter- atomic) QSEI.[33,34] Epitaxial thin films of this material enable control of the magnetic properties through the thickness of the film. We utilized this property to tune the Curie temperature similar to the approach described in Ref. 29. Using this approach, we were able to obtain a film in which we could change the magnetic order during OER by changing the temperature. Moreover, we use the strain induced magnetic anisotropy in the film to further investigate the effect of domain alignment and the magnitude of the magnetization in our films.

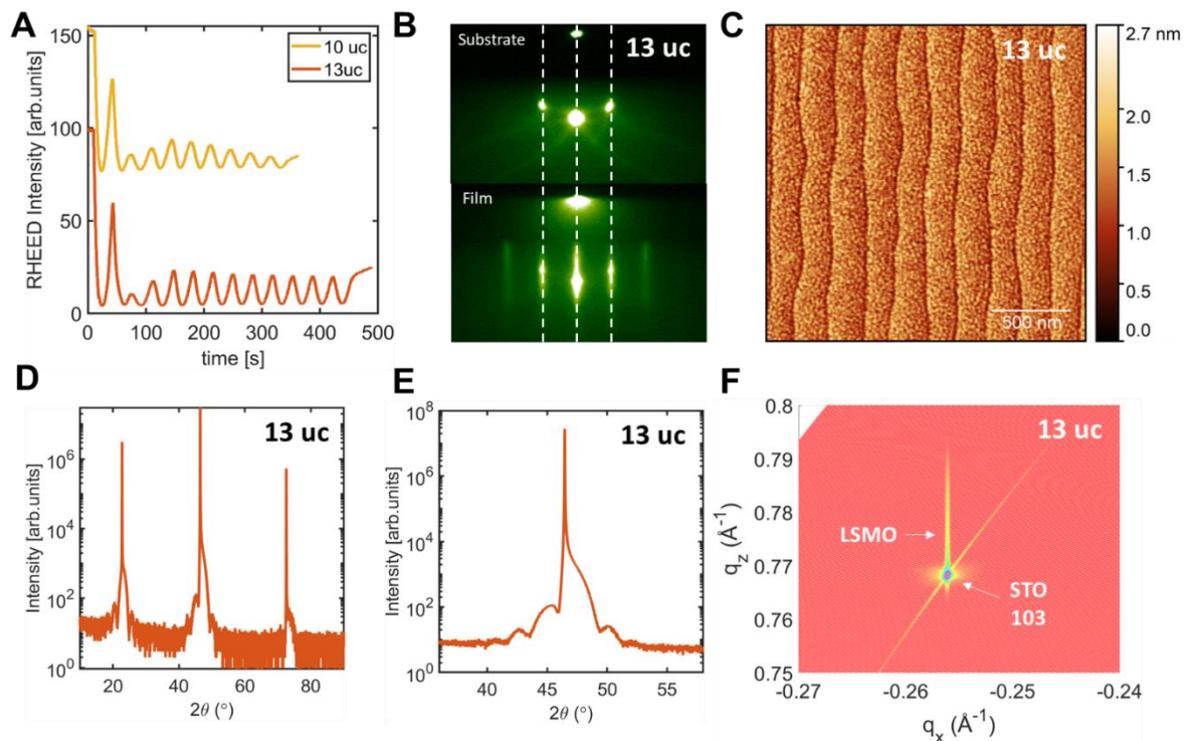

*Figure 1: RHEED and X-ray diffraction of $La_{0.67}Sr_{0.33}MnO_3$ thin films. (a) RHEED intensity during growth of the 10 uc and 13 uc thick films, with clear oscillations indicating a 2D layer-by-layer growth. (b) RHEED diffraction pattern of the substrate and 13 uc thick film. The equal distance between diffraction spots indicates epitaxial growth. The slight elongation of the spots indicates roughening of the film. (c) AFM image of the 13 uc thick film. (d) Wide 2θ-ω scan of the 13 uc thick film which reveals a single phase of 001 $La_{0.67}Sr_{0.33}MnO_3$ (e) 2θ -ω scan of the 002 $SrTiO_3$ peak of the 13 uc thick film with pronounced Laue fringes. (f) Reciprocal space map of the 103 peak of $SrTiO_3$. As the film peak lies along the same value of $q_x$ as the substrate, the film is fully strained on the substrate.*

We synthesized epitaxial thin $La_{0.67}Sr_{0.33}MnO_3$ films of 10- and 13-unit cell (uc) thickness, using PLD parameters similar to Ref. 29. For all films, the growth proceeds in a two-dimensional (2D) layer-by-layer manner, as demonstrated by in-situ reflection high energy electron diffraction (RHEED, Figure 1(a)), and the resulting surface morphologies exhibit the characteristic vicinal step-terraced structure also observed for the substrate surface with small corrugations/islands/decorations on each terrace. (Figure 1(b)). Clear RHEED oscillations during growth indicate the intended thicknesses of 10 uc and 13 uc (Figure 1(a)), further confirmed by X-ray reflectivity (XRR), which revealed thicknesses of ~3.6 nm and ~4.8 nm, respectively (Supplementary figure 3). X-ray diffraction measurements reveal a single La$_{0.67}$Sr$_{0.33}$MnO$_3$ 001 phase (Figure 1(d)). Moreover, the existence of clear Laue fringes indicates a high bulk crystalline quality of the film (Figure 1(e)). Reciprocal space mapping confirms coherent strain to the substrate (Figure 1(f)).

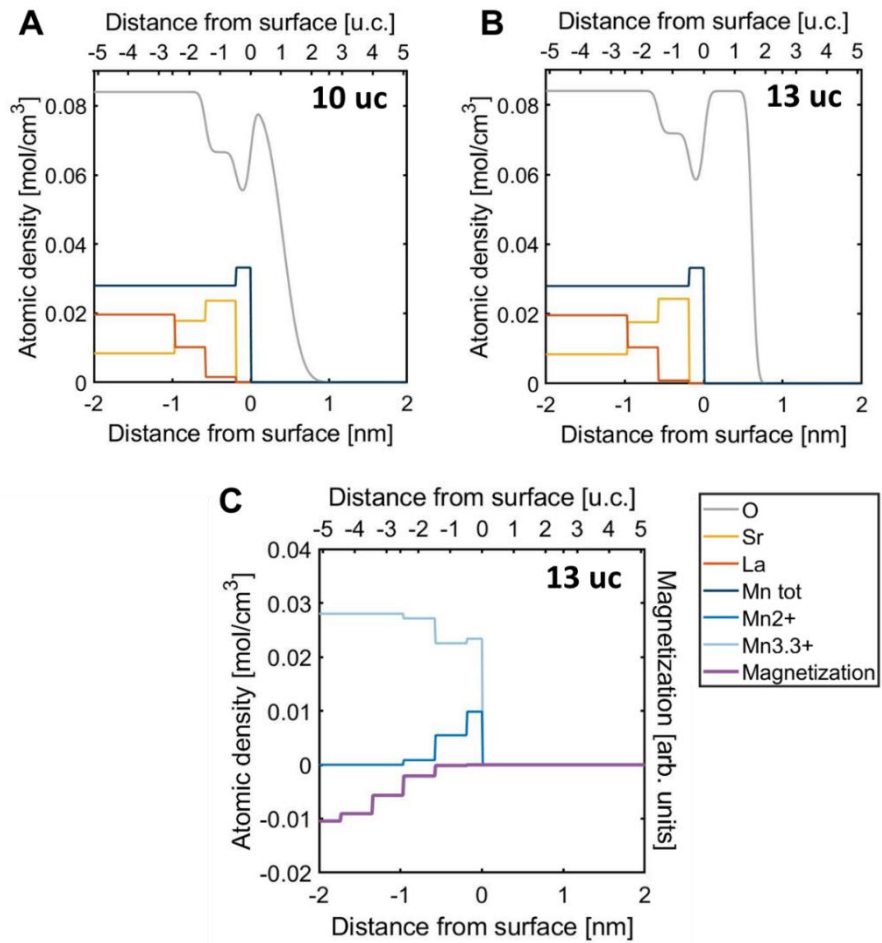

*Figure 2. Unit-cell resolved model obtained from resonant X-ray reflectivity measurements of the films which shows the atomic density of the different elements O, La, Sr and Mn (total or split up between Mn$^{2+}$ and Mn$^{3.3+}$) as a function of the distance from the surface (see methods for further information) for a (a) 10 uc thick La$_{0.67}$Sr$_{0.33}$MnO$_3$ film (b) 13 uc thick La$_{0.67}$Sr$_{0.33}$MnO$_3$ film. Measurements were performed at 150K. (c) Unit-cell resolved model obtained from resonant circular polarized X-ray reflectivity measurements which shows magnetization in the surface region of the film as a function of the distance from the surface. Magnetization is shown alongside the depth dependent atomic density of the Mn species. Magnetization is given in arbitrary units. Magnetization was obtained under a field of 0.6T at 300K and is given in arbitrary units.*

To further investigate the surface structure in our La$_{0.67}$Sr$_{0.33}$MnO$_3$ films, resonant x-ray reflectivity measurements (RXR) were performed. Using this technique, the depth dependence of the atomic densities of each of the components in a thin film can be probed. From the obtained unit-cell resolved stoichiometry of the surface (as shown in Figure 2(a) and (b)), we observe that both films have a similar bulk and surface stoichiometry, both films are B-site terminated and have Sr enrichment and La deficiency at the surface (similar to ref. 35). Moreover, a clear indication for oxygen vacancies at the surface is found, which could be the reason for the Mn$^{2+}$ formation shown in Figure 2(c).

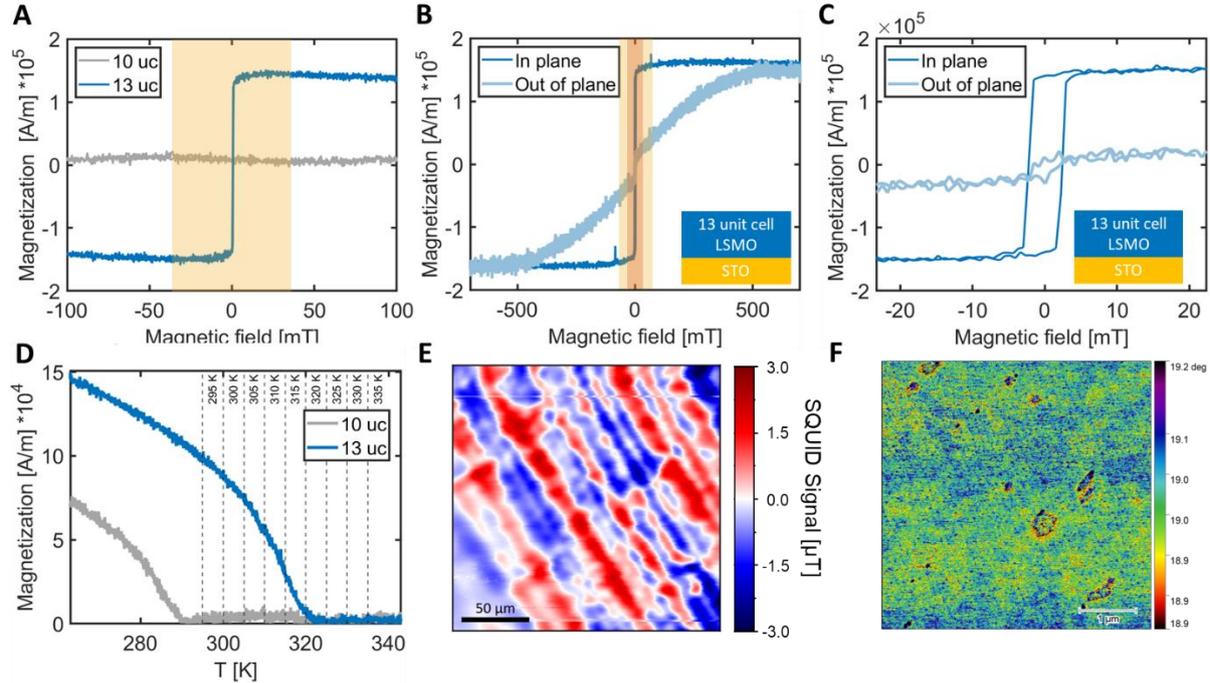

*Figure 3. Magnetic properties of La$_{0.67}$Sr$_{0.33}$MnO$_3$ thin films with a thickness of 10 and 13 uc. (a) In-plane magnetic hysteresis loops measured at 300 K. The diamagnetic contribution to magnetization (not shown) has been attributed to the substrate and has been subtracted. The colored area indicates the in-plane magnetic fields we applied during electrocatalytic experiments described below. (b) In-plane and out-of-plane magnetic hysteresis loops of the 13 uc La$_{0.67}$Sr$_{0.33}$MnO$_3$ film obtained under similar conditions as (a). Colored areas indicate the in-plane (orange) and out-of-plane (yellow) fields we applied during electrocatalytic experiments. (c) Zoom in of the hysteresis loops shown in (b). (d) Temperature dependent magnetization curve measured at 10 mT. All samples were cooled in this field to 20 K before measurement. The dotted lines indicate the temperatures at which electrochemical measurements were performed. (e) Scanning SQUID microscopy measurement performed at 4.2 K on a 13 uc thick film. (f) MFM measurements obtained at room temperature on a 13 uc thick film. The large features are due to crosstalk from the topography shown in Supplementary figure 2. The differences in the background indicate inhomogeneity in the magnetic behavior.*

We explored the magnetic properties of the La$_{0.67}$Sr$_{0.33}$MnO$_3$ films ex-situ using vibrating scanning magnetometry (VSM), scanning superconducting quantum interference device (SQUID) microscopy and magnetic force microscopy (MFM) (Figure 3). We observe room temperature ferromagnetic behavior with a saturation magnetization of 1.6*10$^5$ A/m, equal to approximately 1 $\mu_B$/Mn atom, and a H$_C$ of ~3 mT for the 13 uc film as both a clear hysteresis loop and a $T_c$ of ~322K were found. For the 10 uc film, we observe room temperature paramagnetic behavior because no hysteresis loop is observed due to a $T_c$ of ~290K. We thus successfully prepared two films with different magnetic properties while maintaining crystal structure and stoichiometry in the bulk and only slightly changing the surface chemical composition.

Moreover, as seen in Figure 3(b), the external magnetic response of the ferromagnetic film is anisotropic. A magnetic easy axis is found along the in-plane direction of the film, which was expected due to the tensile strain induced by the substrate.[35] A distinct domain structure with large domains along 1 direction is observed for the 13 uc film at low temperature and zero field as shown in Figure 3(d). However, room temperature MFM measurements reveal a spatial inhomogeneity in the magnetic behavior reflected by the inhomogeneity in the background of Figure 3(f). We hypothesize that this is due to the formation of a mixed phase, consisting of a ferromagnetic-like matrix with a range of different long spin correlation lengths containing local defects. Similar behavior has been reported before for La$_{0.67}$Sr$_{0.33}$MnO$_3$ [36]. These defects may be of a paramagnetic nature. This behavior is further supported by the trend of the resistivity curve with temperature (Supplementary figure 1(b)) which can only be explained if there exists a competition between ferromagnetic (FM) metallic domains and paramagnetic (PM) regions at elevated temperatures, as previously described for similar manganite oxides. [37,38] These paramagnetic domains can be a reason for the low saturation magnetization compared to the low temperature value for bulk $La_{0.67}Sr_{0.33}MnO_3$ of 2.5 $\mu_B$/Mn.

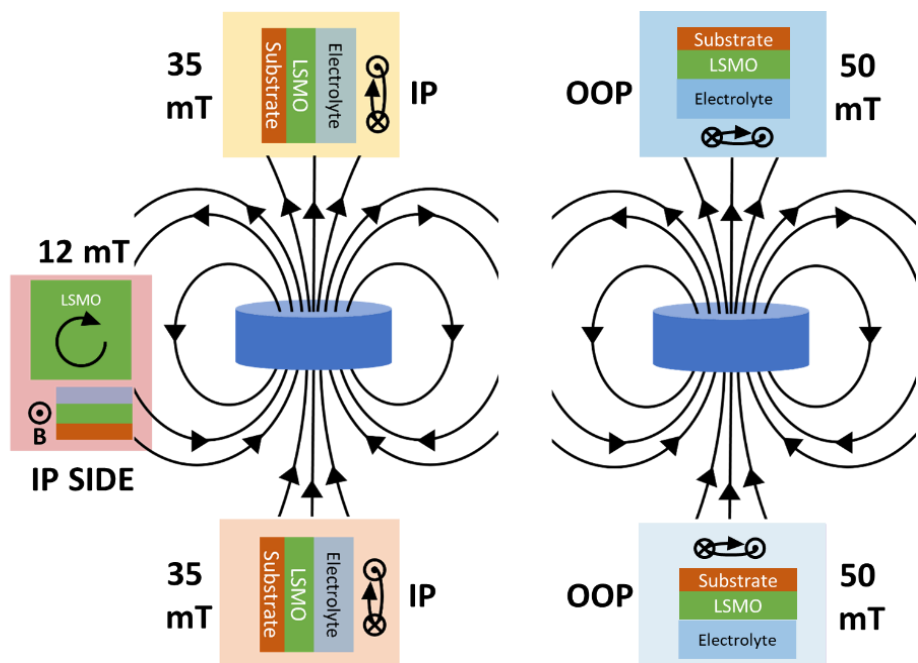

*Figure 4. Schematic of the different field directions applied during external field application experiments. The direction of the rotation of the sample is shown alongside the relative orientation of the sample to the field lines of the magnet where IP and OOP stand for the configuration where the field lines at the sample are respectively parallel or perpendicular to the surface The applied field strengths of 35 mT for the in-plane directions and 50 mT for the out-of-plane directions are indicated for each of the directions. "IP side" is a special case, where the sample was placed at the side of the magnetic instead of being placed at one of the poles (normal IP and OOP case). Here the field strength is only 12 mT at the sample.*

Having established the physical properties of our model catalysts, we now turn to electrochemical investigation. We review the magnetic-field dependent activity first, followed by the assessment of the effect of the intrinsic magnetic order. Cyclic voltammetry measurements were performed, both in the absence and presence of an applied external magnetic field in different in-plane and out-of-plane directions (see Figure 4 and methods for details). The presence (absence) of systematic changes between the activity with and without in-plane applied magnetic field for ferromagnetic (paramagnetic) catalysts indicates that the activity enhancement is linked to the increase in magnetization in the ferromagnetic film upon external field application (Figure 5 (a),(b),(e),(f) and supplementary discussion A. Due to ageing and scattering, the effect of the out-of-plane field exposure could not be distinguished (Figure 5 (c),(d),(g),(h)), a point to which we will return below.

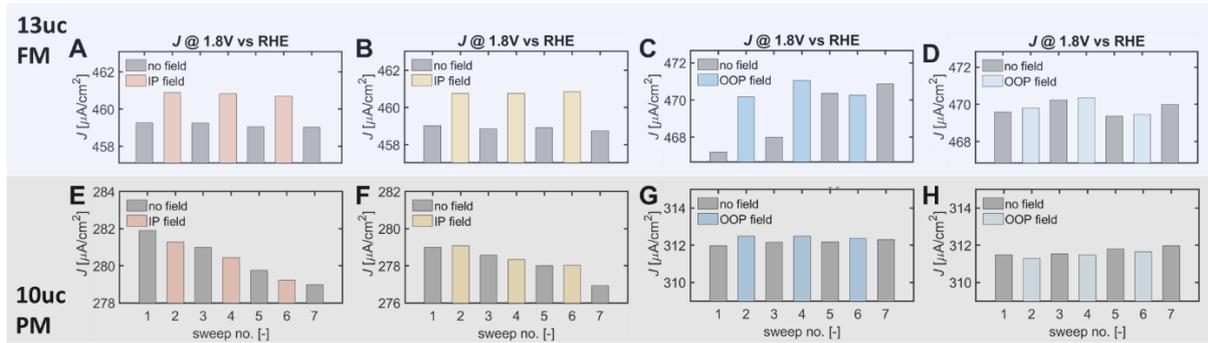

*Figure 5. (a) to (h) Current densities obtained from cyclic voltammetry measurements (see Supplementary figure 4) at an iR-corrected potential of 1.8 V vs RHE for different sweeps. Top row: 13 uc thick $La_{0.67}Sr_{0.33}MnO_3$ film. Bottom row: 10 uc thick $La_{0.67}Sr_{0.33}MnO_3$ film. Colored bars indicate the presence of an external magnetic field during the sweep while grey bars indicate sweeps which were done without the presence of a field. The color of the bar represents the direction of the field and corresponds to the colors indicated in Figure 4.*

To further separate the effects of the external field application from ageing or scattering effects, chronoamperometry measurements were performed at 1.8 V vs RHE. External fields were applied in different in-plane and out-of-plane directions. Again, an enhancement in current density was observed for the ferromagnetic materials upon in-plane field application (Figure 6 (a)). However, for the behavior linked to the out-of-plane fields, the change in current density depends heavily on the direction of the field lines perpendicular to the sample surface (Figure 6 (a) & (c)). Moreover, the paramagnetic sample showed clearly distinguishable changes in current density upon magnetic field exposure in certain directions contrary to previously published results. [6,12,21] This behavior and possible origins are discussed in more detail in supplementary discussion B. In these measurements again ageing effects can be seen as the current density in the periods between magnetic field application slightly increases over time.

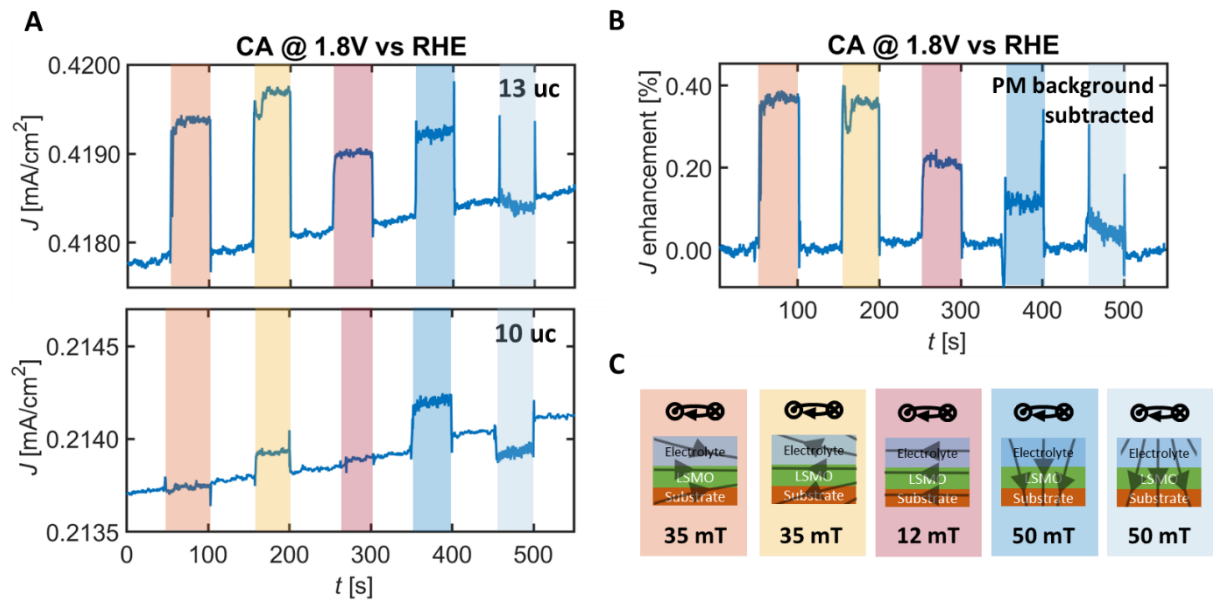

*Figure 6. (a) Chronoamperometry measurement at an iR corrected potential of 1.8V vs RHE for a 13 uc and a 10 uc thick $La_{0.67}Sr_{0.33}MnO_3$ film during which an external magnetic field was applied in different directions. The colored bars indicate the intervals during which the field was applied. The color of the bar represents the direction of the field and corresponds to the colors indicated in Figure 4. (b) Normalized current density enhancement due to ferromagnetic interactions after paramagnetic background subtraction. (c) Schematic drawing of the field lines in the samples in electrolyte. The arrow on top indicates stirring direction.*

To differentiate the effect of the changes in internal magnetic order through domain alignment from other effects also at play for the paramagnetic sample, the paramagnetic background as well as ageing induced effects were subtracted from the response of the ferromagnetic film (Figure 6(b)). Interestingly, an enhancement was found in all field directions for the renormalized data. The current increase is more than two times higher in the in-plane direction than in the out-of-plane direction (~0.4% and 0.1% respectively). This enhancement thus has an anisotropy where a small field along the easy axis has a larger effect on OER activity than a larger field along the hard axis. Comparing the magnitude of the enhancement to the increment in magnetoresistance at these field strengths, which is less than ~0.05% (Supplementary figure 1(a)), we exclude the change in magnetoresistance as a dominant factor for magnetic field enhanced OER activity in this system.

Although we have thus shown that changing the total magnetization in the catalyst can enhance OER activity, we have not yet touched upon the key point of investigating the influence of the intrinsic magnetic order in catalysts without the application of an external field. For this purpose, we explored the Hedvall effect in the OER and changed the intrinsic magnetic order in-situ. We performed OER measurements at different temperatures as indicated in Figure 3(c) (for more information, see methods).

For all samples, a decreasing activity with decreasing temperature was observed (Supplementary figure 5). This loss in current density $J$ is expected because lowering the temperature lowers the thermal energy, lowering the reaction rate of the OER. Simplifying the Butler-Volmer equation using the Tafel approximation at high potentials and neglecting diffusion limitations gives [39]:

$$lnJ = lnJ_o + \alpha_f F\eta \frac{1}{k_B T} \qquad 1$$

for the exchange current density $J_o$, the transfer coefficient of the overall forward reaction $\alpha_f$, Faraday constant $F$, overpotential $\eta$ and Boltzmann constant $k_B$. However, the exact temperature-dependent behavior may deviate from equation 1 because multiple effects like water adsorption and dissociation, the nature of adsorbed species, and movement of oxygen vacancies or interstitials in the catalyst can also be influenced by temperature.[40] While this complicates identification of magnetic-order-effects, a comparison between 10 uc and 13 uc samples enables differentiating different effects, as the chemical composition is similar for both samples and the purely temperature dependent electrochemical effects should be similar. On the contrary, a change in intrinsic magnetic order occurs for the 13 uc thick film in this temperature range ($T_C \approx$ 321 K) while the 10 uc film is paramagnetic over the entire temperature range.

The difference in electrical resistance between these 2 films resulted in differences in absolute activity, necessitating a comparison of the activity trends with varying temperature. We plot $lnJ$ versus $\frac{1}{T}$ for both films in an Arrhenius plot (Figure 7(a)), using offset y-axes to account for the overall activity differences, where the slope is proportional to $\eta$ (eq. 1). For the paramagnetic 10 uc film, we find the expected almost linear decrease of $lnJ$ with $\frac{1}{T}$. But for our ferromagnetic film, there are competing effects for $T<T_c$, reflected by the upwards deviation of the slope of the 13 uc sample for $T<T_c$ (blue region) compared to the paramagnetic 10 uc sample, whereas for $T>T_c$ (yellow region), the slopes of both samples are similar (Figure 7(a)). The positive change in slope implies a decreased overpotential or a change in transfer coefficient, increasing the OER activity. We observe an enhancement of ~35% compared to the expected current density without ferromagnetic order (see Supplementary figure 6 for derivation). As discussed in supplementary discussion C, sample degradation, surface restructuring and changes in resistance with temperature cannot explain the observed effects. We can thus conclude that the occurrence and continuous increase of the ferromagnetic order with decreasing temperature induces the relative current density enhancement below $T_C$.

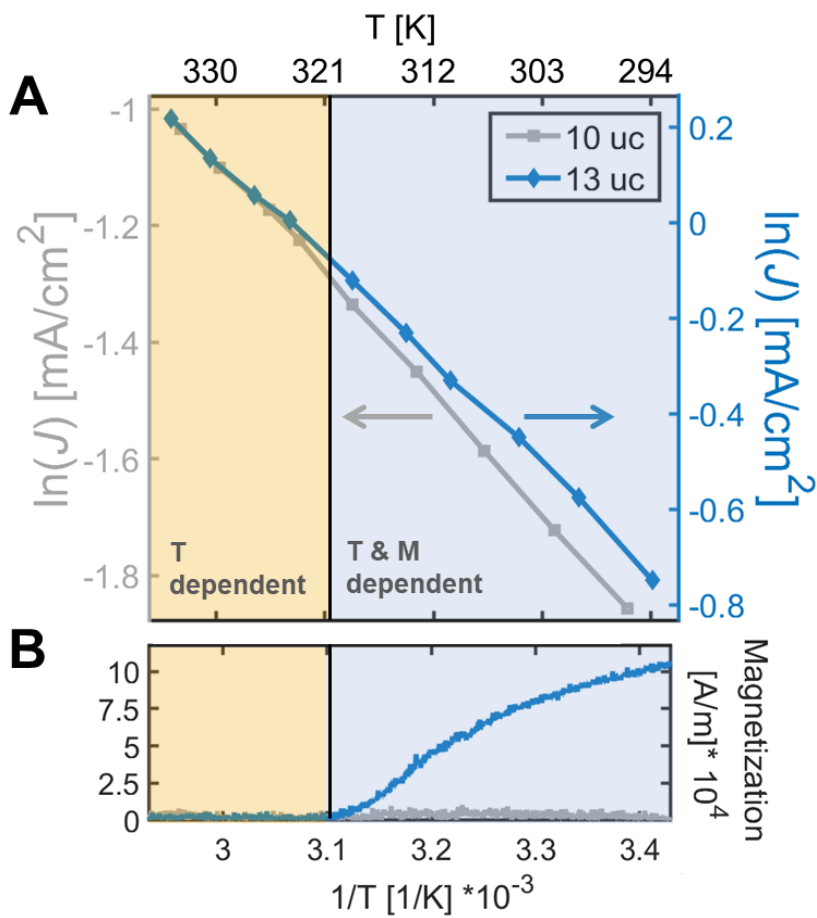

*Figure 7 (a) Plot of the natural logarithm of the current density versus the inverse of the temperature obtained for a ferromagnetic (13 uc) and a paramagnetic (10 uc) $La_{0.67}Sr_{0.33}MnO_3$ film at an iR corrected potential of 1.8 V vs RHE. The plot is shown alongside the magnetization (b) of the films in the same temperature range.*

To summarize, we have shown a relative current density increase during OER upon changing the magnetic properties of $La_{0.67}Sr_{0.33}MnO_3$ thin film catalysts. The combination of effects shown in the temperature-dependent OER activity measurements for a ferromagnetic catalyst and the effects shown while exposing the same ferromagnetic catalysts to an external magnetic field during OER validate that the enhancement is primarily induced by the changes in magnetic order in the catalyst.

The comparably small enhancement (see Supplementary table 1 for a detailed comparison to prior works) may be explained by the small applied external field, small coercivity and low saturation magnetization of the $La_{0.67}Sr_{0.33}MnO_3$ films at 300K. Most importantly, the difference between saturation and initial magnetization in $La_{0.67}Sr_{0.33}MnO_3$ films is comparatively small due to a small coercive field, large domains and the presence of paramagnetic regions which lower the increment in magnetization upon alignment as discussed in depth by Jingjie Ge et al.[14]. Moreover, both the effects of domain alignment and a change in magnetic order could be influenced by the existence of a magnetic dead-layer, in which any long-range magnetic order is diminished. Our epitaxial model system is a suitable platform to further investigate the effect of such dead-layers, similar to previous studies focusing on the surface properties of $La_{0.67}Sr_{0.33}MnO_3$ films.[41–43]

To further investigate where the magnetic order resides in our $La_{0.67}Sr_{0.33}MnO_3$ films, we again use RXR in vacuum. Using left- and right-circularly polarized light, a depth profile of the magnetization can be obtained from the best fit of the asymmetric spectra (Figure 2 (c)), for more information see methods).[44,45] We found that the net magnetization in the top 1.5 uc of the 13 uc thick $La_{0.67}Sr_{0.33}MnO_3$ film is zero and that the magnetization is quenched in the subsurface layer up to 4 uc, even under an applied field. This decreased magnetization is likely connected to the off-stoichiometry discussed above. As these off-stoichiometries likely deteriorate long-range ordered magnetic states, we hypothesize that the 1.5 uc thick surface layer is paramagnetic. Although these measurements were done ex-situ, the paramagnetic surface layer is likely also present under OER conditions, because the fields applied in the electrochemical cell are smaller than the fields during RXR measurements.

The lack of lateral long-range magnetic order in the catalyst surface implies that no ordered interaction exists between the long-range ordered spin states of the metal atoms in the film and the reaction intermediates. We can thus conclude that the observed OER enhancements due to inter-atomic QSEI are mainly induced by the magnetic order in the bulk of the $La_{0.67}Sr_{0.33}MnO_3$ films. Still the intra-atomic QSEI is always present and persists in the surface layer. The spin-polarization of the adsorbents is thus presumably mediated through the out-of-plane QSEI in the paramagnetic surface layer, which is smaller than the ferromagnetic QSEI in the bulk, limiting the efficiency of the OER enhancement.

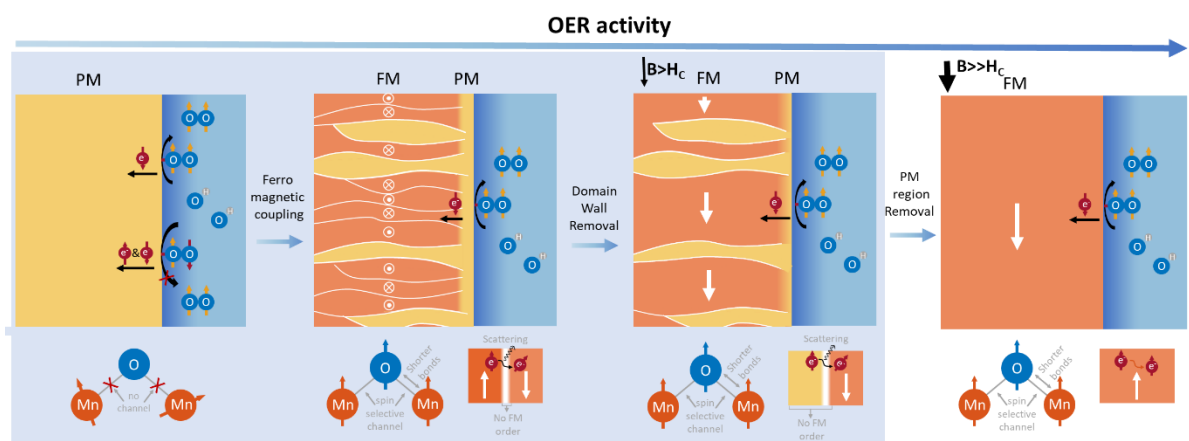

*Figure 8. Proposed mechanism for spin-polarized OER activity enhancement in epitaxial $La_{0.67}Sr_{0.33}MnO_3$ thin films. In the left panel we show a fully paramagnetic catalyst layer which corresponds to our 10 uc sample for the entire T range investigated here and the 13 uc sample at $T>T_C$. In the second panel, the catalytic film is made up of a ferromagnetic matrix with domains in which paramagnetic regions are embedded. Moreover, a paramagnetic surface layer is present. This state corresponds to the intrinsic magnetic state of our 13 uc film at $T<T_C$. The third panel shows the state of the catalyst after domain removal in the ferromagnetic matrix upon external magnetic field exposure. The right panel shows the ideal state of the 13 uc film after removal of the paramagnetic regions and surface layer, a homogeneous ferromagnetic catalyst. The states achieved in this work are highlighted with a blue square. At the bottom of the image, we show the physical interactions in the film which account for the presence/absence of spin selectivity during adsorption of electrons. The reaction intermediates shown are highly simplified to focus only on the formation of the O-O bond which is the most important bond when considering the spin-polarized OER mechanism as the spin alignment in this bond either blocks or allows the formation of triplet oxygen.[17]*

Based on the findings from the epitaxial $La_{0.67}Sr_{0.33}MnO_3$ thin film model system and considering the range of observations in recent literature, we propose a unifying picture for the spin-polarized OER activity enhancement oxide catalysts, schematically shown in Figure 8. The lowest activity is found for a paramagnetic material with no spin channels (left panel).[7,30] Because of the absence of stable spin channels in this catalyst, O-O intermediates with parallel and antiparallel spin alignment can both be formed. The latter do not have optimum inter-site quantum spin exchange interactions and electronic quantum correlations are not optimum for the thermodynamic overpotential.

If ferromagnetic ordering comes into play below $T_c$, a film consisting of a ferromagnetic bulk with small embedded paramagnetic regions and a paramagnetic surface is obtained (Figure 8, second panel). The ferromagnetic exchange interactions in the matrix induce ordered itinerant spin channels, inducing spin-selective electron mobility. This leads to a higher possibility of the formation of bonded spin ordered O-O intermediates which in turn increases the OER activity. [7,10] Moreover, the QSEI associated with the spin orbital ordering lowers the interatomic repulsions, which lead to shorter metal-oxygen bonds, which may act favorably on the reaction intermediate binding energies.[10] All of these aspects facilitate the generation of triplet $O_2$ molecules for ferromagnetic thin film catalysts. [10] However, the described ferromagnetic film still contains domains and domain walls. The domain walls induce electron scattering, effectively lowering spin selective transport. [11,15] Moreover, the presence of domain walls and paramagnetic regions lowers the amount of ferromagnetic reaction sites [15] and the paramagnetic surface lowers the effectiveness of the interaction between the ferromagnetic domains in the bulk of the film and the adsorbents.

Upon domain wall removal under the influence of an external magnetic field a uniform ferromagnetic matrix is obtained (Figure 8, third panel). The removal of domains lowers the amount of domain wall scattering and increases the amount of long-range magnetically ordered reaction sites. Moreover, the exchange field between ferromagnetic bulk and paramagnetic surface layer becomes stronger, which may become thinner as a result, enhancing the interaction between the ferromagnetic layer and the adsorbents. [17] The effectiveness of domain alignment depends on the strength and direction of the magnetic field. However, as paramagnetic regions remain, OER activity is still not optimal. The application of higher external fields could further improve the performance, by increasing the net magnetic moment in the paramagnetic regions until saturation is achieved and the paramagnetic regions and surface layer are completely removed (Figure 8, right panel). [6,16] This effect could explain the differences in activity enhancement between this work and the results shown in Ref. 19, where much higher fields were applied. This fully ferromagnetic state is not accessible in our model system using realistic fields. Further research thus needs to be done to verify OER enhancement in such a fully long-range ordered electrocatalyst.

**CONCLUSION**

In conclusion, we employed epitaxial $La_{0.67}Sr_{0.33}MnO_3$ thin film model catalysts to arrive at a unifying picture of the effects of intrinsic magnetic order and applied magnetic fields on the oxygen evolution reaction. We tuned the magnetic order in the films in-situ during OER by exploiting the para- to ferromagnetic transition at $T_c$. Using this strategy, we showed that the presence of ferromagnetic ordering below the Curie temperature enhances OER activity. In the ferromagnetic films, application of external magnetic fields is linked to a further increase in OER activity. Moreover, we observed a correlation between the magnetic anisotropy in our catalyst and the external magnetic field induced OER enhancement. Our work thus suggests that both the intrinsic magnetic order in $La_{0.67}Sr_{0.33}MnO_3$ films and externally triggered changes in the magnetic structure affect the catalytic activity of these films. The OER enhancements due to inter-atomic QSEI are found to primarily result from changes in the magnetic order of the bulk of the catalyst, because no long-range magnetic ordering existed at the top of the catalytic surface. To further verify the proposed unifying picture, further research could focus on operando magnetic characterization to directly correlate the magnetic order and interactions in these catalysts with the activity under realistic OER conditions.

**SUPPLEMENTARY MATERIAL**

See the supplementary material for more information that supports the findings of this work: Additional electrochemical data used to generate the overview in the main text. Schematics and images of the experimental setups. Additional thin film characterization, both before and after electrochemical tests. Additional discussion of results.

**ACKNOWLEDGMENTS**

Support from the University of Twente in the framework of the tenure track start-up package is gratefully acknowledged. Robert J. Green acknowledges support from the Natural Sciences and Engineering Research Council of Canada (NSER(c) Discovery Grant program. Lucas Korol acknowledges support from the NSERC CREATE to INSPIRE program. Magnetocat acknowledges the funding from the European Union's, Horizon 2020 research and innovation program ,under grant agreement No. 964972 (H2020-FETOPEN-2018-2019-2020-01). Chiara Biz, Mauro Fianchini and Jose Gracia thank the SpinCat consortium.

**AUTHOR DECLARATIONS**

**Conflict of Interest**

The authors have no conflicts to disclose.

**Author Contributions**

**E. van der Minne:** Conceptualization (lead); Methodology (lead); Data curation (lead); Investigation (lead); Writing – original draft (lead); Writing – review and editing (lead); Validation (lead); Visualization (lead); **L. Korol:** Investigation RXR (lead); Analysis RXR (lead); **L.A.M. Krakers:** Investigation electrochemistry under field (support); **M. Verhage:** Investigation MFM (lead); Analysis MFM (lead); Writing – review and editing (support); **C.M.M. Rosário:** Investigation SQUID (lead); Analysis SQUID (lead); **T.J. Roskamp:** Investigation SQUID (lead); Analysis SQUID (lead); **R.J. Spiteri**: Software (support); Writing — review and editing (support) **C. Biz**: Methodology (support); Writing – review and editing (support); **M. Fianchini**: Methodology (support); Writing – review and editing (support); **G. Rijnders**: Methodology (support); Writing – review and editing (support); **K. Flipse**: Methodology (support); Writing – review and editing (support); **J. Gracia**: Theoretical understanding (lead); Writing – review and editing (support); **G. Mul**: Conceptualization (support); Methodology (support); Writing – review and editing (support); **R.J. Green:** Investigation RXR (lead); Analysis RXR (lead); **H. Hilgenkamp**: Conceptualization (support); Methodology (support); Writing – review and editing (support); Analysis SQUID (support); **G. Koster**: Supervision (support); Conceptualization (support); Methodology (support); Writing – review and editing (support); **C. Baeumer:** Conceptualization (lead); Methodology (lead); Data curation (lead); Investigation (supporting); Supervision (lead); Writing – original draft (support); Writing – review and editing (lead); Validation (support)

**DATA AVAILABILITY**

The data that support the findings of this study are available within the article and the supplementary material. Additional data is available from the corresponding authors upon reasonable request.

## METHODS

**Pulsed Laser Deposition**

$La_{0.67}Sr_{0.33}MnO_3$ thin films are deposited via reflection high-energy electron diffraction (RHEE(d)-assisted pulsed laser deposition (PL(d)) onto B-site terminated and step-terraced $SrTiO_3$ (001) substrates purchased from CrysTec GmbH or Shinkosha Co., Ltd. A stochiometric $La_{0.67}Sr_{0.33}MnO_3$ target was obtained from SurfaceNet. The films were deposited with a laser fluence of 2.0 J cm$^{-2}$ and a frequency of 1 Hz. The deposition was done at an oxygen pressure of 0.266 mbar, and the temperature of the substrate was kept at 750 °C during deposition. The distance between sample and target was kept at 5 cm, and a rectangular mask was used to obtain a laser spot size of 2.24 mm$^2$. Before deposition, the targets were pre-ablated at 5 Hz. After deposition, the samples were cooled down at 25 °C/min inside the PLD at 100 mbar oxygen pressure. PLD was performed in a vacuum system (TSST) with a base pressure of $5 \times 10^{-8}$ mbar, equipped with an in-situ RHEED (Staib instruments) and a KrF excimer laser (Coherent, Inc.) of 248 nm.

**Thin film characterization**

X-ray diffraction and reflectivity measurements were performed using a Panalytical X'pert pro diffractometer with Cu anode. For the diffractograms, a GE 220 Monochromator was used to obtain Cu-K$_\alpha$ radiation. During the 2Θ-ω scans, the detector was operated in 0D mode with an active length of 0.165 mm. A slit of 1/2º was used to shape the beam. Reflectivity measurements were performed using a of 1/32º slit. The detector was operated in 0D mode with an active length of 1.12 mm. The RSM was performed using a Bruker D8 Discover diffractometer with Cu-K$_\alpha$ radiation and an Eiger 2 R 500K area detector. The detector was kept stationary while operated in 1D mode as an omega rocking curve was performed. A grazing-exit configuration was chosen to obtain narrow diffactograms.

The topography of the grown films was characterized by atomic force microscopy (AFM) using a Veeco Dimension Icon AFM in tapping mode in air. The oscillating cantilever is a Tespa-V2 cantilever (Bruker, Netherlands) with a pure silicon tip with a nominal radius of 20 nm. Images were obtained using the Nanoscope software and treated using the Gwyddion software. First, we align the rows using a polynomial approach, then the data is leveled by mean plane subtraction, lastly a polynomial background was subtracted.

**Magnetic characterization**

VSM measurements were obtained using a DYNACOOL physical properties measurement system (Quantum Design, Germany). All samples were field cooled at 10 mT before measuring temperature-dependent magnetization.

Scanning superconducting-quantum-interference-device (SQUI(d)) microscopy measurements were performed at 4.2 K in a liquid He bath. A superconducting Nb shield is used to shield the sample and SQUID sensor from any external magnetic fields. The SQUID sensor used is extended with a pickup loop having an effective area of approximately 18 μm$^2$, and scanning is performed under an angle of approximately 15º, making the distance between the pickup loop and sample surface approximately 2-3 μm. The SQUID is a flux-to-voltage transducer, and the measured voltage can be converted to magnetic field by dividing by the (measurement-dependent) flux-to-voltage ratio and the effective area of the pickup loop. Typical flux sensitivities are in the order of 10-20 μPhi$_0$/(Hz)$^{1/2}$, and the bandwidth is 1000 Hz. [46,47]

Magnetic force microscopy was performed with a CoCr coated Si cantilever, with a spring constant of 2.5 N/m and a resonant frequency around 65 kHz. Prior to imaging a permanent magnet was used to magnetize the tip for sensitivity of in-plane magnetic fields. A Veeco Dimension AFM III was operated in tapping mode feedback. Image was obtained at 1Hz scan speed with 512 pixels x 512 lines. The magnetic signal was obtained by lift mode, with a lift height of 10 nm. Gywddion was used for image processing with plane subtraction and line alignment via means of differences.

**Electrochemical characterization**

To perform electrochemical experiments with epitaxial thin films on 10 × 10 × 0.5 mm single crystal substrates, we used a custom-made adapter to press the sample back side to the Pt plug of a rotating disk electrode (RDE, Pine Research). For further details, see Supplementary figure 7. 50 nm Pt connections from the sample back side to the front side ensured electrical contact with the Nb:$SrTiO_3$ substrate and the epitaxial layers. On the front side, a film area of 7.5 mm diameter was exposed to the electrolyte and sealed using an O-ring (Kalrez, ERIKS, Germany). The RDE shaft was rotated at 1600 rpm. Electrochemical testing was performed using a Biologic SP-300 potentiostat in a 150 mL alkaline-resistant cup with a Pt wire as a counter electrode. Electrochemical impedance spectroscopy was conducted with the amplitude of 10 mV at open-circuit potential, and the correction for the cell resistance (IR correction,) was based on the high-frequency intercept of the real impedance. The electrolyte solution of 0.1 M KOH was prepared by dissolving KOH pellets (Sigma-Aldrich, 99.99%) in Milli-Q water. The electrolyte was $O_2$-saturated prior to testing for at least 30 min and maintained under an $O_2$ atmosphere during testing. All electrochemical measurements were performed at room temperature for the domain alignment tests. For elevated temperature experiments, the electrochemical setup was placed in a heater water bath (see Supplementary figure 7). The temperature of the solution was measured in close approximation to the working electrode to determine the temperature near the working electrode. After heating to each temperature, we waited for approximately 20 minutes to ensure temperature stabilization. Potentials were referenced to a Hg/HgO reference electrode (C3 Prozess-und Analysentechnik, Germany). The potential of the Hg/HgO was measured against a reversible hydrogen electrode (RH(e) (HydroFlex, US(a) in 0.1 M KOH. The applied potentials were converted to RHE scale using the measured difference and were corrected using iR correction [48]. All of the OER testing was performed on a fresh electrode that had not undergone previous testing. Cyclic voltammetry was first performed in the pseudocapacitive redox phase change region (~1.1 to 1.5 V vs RH(e) at scan rates between 10 and 500 mV s$^{-1}$, followed by OER testing performed from 1.1 V to 1.9 V vs RHE at a scan rate of 10 mV s$^{-1}$. External magnetic fields were applied using a 1 T permanent disk magnet such that the applied field strength depended on the distance between the magnet and the sample. These magnetic field strengths were measured as a function of distance for each of the directions using a flat hall probe.

**RXR**

The Resonant x-ray reflectometry (RXR) data was acquired at the Resonant Elastic and Inelastic X-ray Scattering (REIXS) beamline of the Canadian Light Source (CLS) in Saskatoon, Canada [49]. The beamline has a flux of $5 \times 10^{12}$ photons $s^{-1}$ and has a photon energy resolution $\Delta E/E$ ~$10^{-4}$. An elliptically polarizing undulator (EPU) is used to give the desired linear and circular polarizations. The experimental chamber was kept below a base pressure of $10^{-9}$ torr, and the measurements were taken at a temperature of 300 K. The samples were aligned with their surface normal in the scattering plane, and the reflection-geometry scans were aided by the in-vacuum 4-circle diffractometer. The measurements were performed in the specular reflection geometry with several resonant photon energies at Ti $L_{2,3}$ (~450-470 eV), Mn $L_{2,3}$ (~635-660 eV), and La $M_{4,5}$ (~830-860 eV) resonances, along with multiple non-resonant photon energies. The circular-dichroic magnetic measurements were performed by inserting a permanent magnet array into the sample environment producing a homogenous 0.6 Tesla field, aligning the magnetization in both the film $xy$ plane and the measurement scattering plane. The dichroic measurements were then taken solely using the resonant photon energies of the Mn $L_{2,3}$ resonance. A photodiode was used to detect the reflected beam intensity, with the response function of the photodiode determined by directly measuring the synchrotron beam. The measured data was normalized by the incident beam flux and the response function to obtain a quantitative reflectivity spectra.

The modelling of the RXR data was performed using the Global Optimization of Resonant X-ray Reflectometry (GO-RXR), a software package recently developed by the qMaX group at the University of Saskatchewan. Tabulated atomic form factors were used for non-resonant energies, while the resonant scattering tensors for elements Ti, Mn, and La were constructed using measured x-ray absorption spectra. For Mn, two different resonant scattering tensors were used – one for $Mn^{2+}$ and one for Mn in stoichiometric $La_{0.7}Sr_{0.3}MnO_3$ (implemented as a weighted linear combination of $Mn^{3+}$ and $Mn^{4+}$ scattering tensors corresponding to $Mn^{3.3+}$). To determine the optical and magneto-optical profiles, a slab model was used that is made up of parametrized layers with defined elements, oxidation states, thicknesses, densities, and roughness. The model parameters are then used to construct an element-specific continuous depth-dependent density profile. The density profile, along with the form factors, are then used to determine the energy- and depth-dependent optical profile. The optical profile is then used to simulate the reflectivity for a given energy, reflection angle, and polarization. To determine the density profile of the 10uc and 13uc $La_{0.67}Sr_{0.33}MnO_3$ samples, the parameters of layer thickness, density, roughness, and Mn oxidation state were optimized while fitting the simulation to the experimental data. To reduce the parameter set, the concentration ratio of Sr and La was fixed to 3:7 throughout the bulk of the film (from the target stoichiometry), but it was allowed to vary near the interface and surface. To determine the magnetic profile of Mn, the nonmagnetic elemental density profile was determined first by optimizing the parameters against an extended sigma-polarized experimental dataset. The magnetic density profile was then fit to the circular polarized data using the asymmetry of the right $R_R$ and left $R_L$ circular polarizations where $A = (R_L - R_R)/(R_L + R_R)$. We show the experimental RXR data and associated fits of the resonant and non- resonant theta/two-theta reflectivity scans and circular polarized asymmetry scans at different energies in Supplementary figure 11 and Supplementary figure 12. Moreover the Mn-resonant energy scans and Mn-resonant circular polarized asymmetry energy scans along with associated fits are displayed in Supplementary figure 13.

**REFERENCES**

[1] G.W. Crabtree, and M.S. Dresselhaus, "The hydrogen fuel alternative," MRS Bull **33**(4), 421–428 (2008).

[2] Z. Abdin, A. Zafaranloo, A. Rafiee, W. Mérida, W. Lipiński, and K.R. Khalilpour, "Hydrogen as an energy vector," Renewable and Sustainable Energy Reviews **120**, (2020).

[3] D. Antipin, and M. Risch, "Trends of epitaxial perovskite oxide films catalyzing the oxygen evolution reaction in alkaline media," JPhys Energy **2**(3), (2020).

[4] Y. Sun, S. Sun, H. Yang, S. Xi, J. Gracia, and Z.J. Xu, "Spin-Related Electron Transfer and Orbital Interactions in Oxygen Electrocatalysis," Advanced Materials **32**(39), (2020).

[5] J. Yao, W. Huang, W. Fang, M. Kuang, N. Jia, H. Ren, D. Liu, C. Lv, C. Liu, J. Xu, and Q. Yan, "Promoting Electrocatalytic Hydrogen Evolution Reaction and Oxygen Evolution Reaction by Fields: Effects of Electric Field, Magnetic Field, Strain, and Light," Small Methods **4**(10), (2020).

[6] X. Ren, T. Wu, Y. Sun, Y. Li, G. Xian, X. Liu, C. Shen, J. Gracia, H.J. Gao, H. Yang, and Z.J. Xu, "Spin-polarized oxygen evolution reaction under magnetic field," Nat Commun **12**(1), (2021).

[7] J. Gracia, R. Sharpe, and J. Munarriz, "Principles determining the activity of magnetic oxides for electron transfer reactions," J Catal **361**, 331–338 (2018).

[8] O. V. Khavryuchenko, V.D. Khavryuchenko, and D. Su, "Spin catalysts: A quantum trigger for chemical reactions," Chinese Journal of Catalysis **36**(10), 1656–1661 (2015).

[9] R.R. Nazmutdinov, E. Santos, and W. Schmickler, "Spin effects in oxygen electrocatalysis: A discussion," Electrochem Commun **33**, 14–17 (2013).

[10] C. Biz, M. Fianchini, and J. Gracia, "Strongly Correlated Electrons in Catalysis: Focus on Quantum Exchange," ACS Catal **11**(22), 14249–14261 (2021).

[11] J. Zou, M. Zheng, Z. Li, X. Zeng, and J. Huang, "Magnetization triggered oxygen evolution reaction enhancement for ferromagnetic materials," Journal of Materials Science: Materials in Electronics **33**(9), 6700–6709 (2022).

[12] Y. Zhang, P. Guo, S. Li, J. Sun, W. Wang, B. Song, X. Yang, X. Wang, Z. Jiang, G. Wu, and P. Xu, "Magnetic field assisted electrocatalytic oxygen evolution reaction of nickel-based materials," J Mater Chem A Mater **10**(4), 1760–1767 (2022).

[13] F.A. Garcés-Pineda, M. Blasco-Ahicart, D. Nieto-Castro, N. López, and J.R. Galán-Mascarós, "Direct magnetic enhancement of electrocatalytic water oxidation in alkaline media," Nat Energy **4**(6), 519–525 (2019).

[14] J. Ge, X. Ren, R.R. Chen, Y. Sun, T. Wu, S.J.H. Ong, and Z.J. Xu, "Multi-Domain versus Single-Domain: A Magnetic Field is Not a Must for Promoting Spin-Polarized Water Oxidation," Angewandte Chemie - International Edition, (2023).

[15] X. Ren, T. Wu, Z. Gong, L. Pan, J. Meng, H. Yang, F.B. Dagbjartsdottir, A. Fisher, H.J. Gao, and Z.J. Xu, "The origin of magnetization-caused increment in water oxidation," Nat Commun **14**(1), 2482 (2023).

[16] T. Sun, Z. Tang, W. Zang, Z. Li, J. Li, Z. Li, L. Cao, J.S. Dominic Rodriguez, C.O.M. Mariano, H. Xu, P. Lyu, X. Hai, H. Lin, X. Sheng, J. Shi, Y. Zheng, Y.-R. Lu, Q. He, J. Chen, K.S. Novoselov, C.-H. Chuang, S.

Xi, X. Luo, and J. Lu, "Ferromagnetic single-atom spin catalyst for boosting water splitting," Nat Nanotechnol, (2023).

[17] T. Wu, X. Ren, Y. Sun, S. Sun, G. Xian, G.G. Scherer, A.C. Fisher, D. Mandler, J.W. Ager, A. Grimaud, J. Wang, C. Shen, H. Yang, J. Gracia, H.J. Gao, and Z.J. Xu, "Spin pinning effect to reconstructed oxyhydroxide layer on ferromagnetic oxides for enhanced water oxidation," Nat Commun **12**(1), (2021).

[18] J. Ge, R.R. Chen, X. Ren, J. Liu, S.J.H. Ong, and Z.J. Xu, "Ferromagnetic–Antiferromagnetic Coupling Core–Shell Nanoparticles with Spin Conservation for Water Oxidation," Advanced Materials **33**(42), (2021).

[19] H. Xu, J. Qi, Y. Zhang, L. Hu, M. Feng, and W. Lü, "Enhanced oxygen evolution reaction *via* the tunability of spin polarization and electronic states in a flexible van der Waals membranous catalyst," Physical Chemistry Chemical Physics **25**(30), 20259–20266 (2023).

[20] H. Xu, J. Qi, Y. Zhang, H. Liu, L. Hu, M. Feng, and W. Lü, "Magnetic Field-Enhanced Oxygen Evolution Reaction via the Tuneability of Spin Polarization in a Half-Metal Catalyst," ACS Appl Mater Interfaces, (2023).

[21] V. Gatard, J. Deseure, and M. Chatenet, "Use of magnetic fields in electrochemistry: A selected review," Curr Opin Electrochem **23**, 96–105 (2020).

[22] L. Lin, R. Xin, M. Yuan, T. Wang, J. Li, Y. Xu, X. Xu, M. Li, Y. Du, J. Wang, S. Wang, F. Jiang, W. Wu, C. Lu, B. Huang, Z. Sun, J. Liu, J. He, and G. Sun, "Revealing Spin Magnetic Effect of Iron-Group Layered Double Hydroxides with Enhanced Oxygen Catalysis," ACS Catal **13**(2), 1431–1440 (2023).

[23] Y. Ma, T. Wang, X. Sun, Y. Yao, H. Chen, G. Wu, C. Zhang, and Y. Qin, "Enhanced Oxygen Evolution of a Magnetic Catalyst by Regulating Intrinsic Magnetism," ACS Appl Mater Interfaces **15**(6), 7978–7986 (2023).

[24] R.R. Chen, Y. Sun, S.J.H. Ong, S. Xi, Y. Du, C. Liu, O. Lev, and Z.J. Xu, "Antiferromagnetic Inverse Spinel Oxide $LiCoVO_4$ with Spin-Polarized Channels for Water Oxidation," Advanced Materials **32**(10), (2020).

[25] P. Huang, M. Meng, G. Zhou, P. Wang, W. Wei, H. Li, R. Huang, F. Liu, and L. Liu, "Dynamic orbital hybridization triggered spin-disorder renormalization via super-exchange interaction for oxygen evolution reaction," PNAS **120**(21), (2023).

[26] Z. Sun, L. Lin, J. He, D. Ding, T. Wang, J. Li, M. Li, Y. Liu, Y. Li, M. Yuan, B. Huang, H. Li, and G. Sun, "Regulating the Spin State of $Fe^{III}$ Enhances the Magnetic Effect of the Molecular Catalysis Mechanism," J Am Chem Soc, (2022).

[27] Y. Tong, Y. Guo, P. Chen, H. Liu, M. Zhang, L. Zhang, W. Yan, W. Chu, C. Wu, and Y. Xie, "Spin-State Regulation of Perovskite Cobaltite to Realize Enhanced Oxygen Evolution Activity," Chem **3**(5), 812–821 (2017).

[28] J. Pan, T. Li, and Y. Shan, "Spin-Dependent Reconstruction Induced by Surface Symmetry Breaking in Manganese Spinel Oxides Toward Acidic Oxygen Evolution Reaction," Physica Status Solidi - Rapid Research Letters, (2023).

[29] M. Huijben, L.W. Martin, Y.H. Chu, M.B. Holcomb, P. Yu, G. Rijnders, D.H.A. Blank, and R. Ramesh, "Critical thickness and orbital ordering in ultrathin $La_{0.7}Sr_{0.3}MnO_3$ films," Phys Rev B **78**(9), (2008).

[30] C. Biz, J. Gracia, and M. Fianchini, "Review on Magnetism in Catalysis: From Theory to PEMFC Applications of 3d Metal Pt-Based Alloys," Int J Mol Sci **23**(23), (2022).

[31] M.L. Weber, and F. Gunkel, "Epitaxial catalysts for oxygen evolution reaction: Model systems and beyond," JPhys Energy **1**(3), (2019).

[32] C. Baeumer, J. Li, Q. Lu, A.Y.L. Liang, L. Jin, H.P. Martins, T. Duchoň, M. Glöß, S.M. Gericke, M.A. Wohlgemuth, M. Giesen, E.E. Penn, R. Dittmann, F. Gunkel, R. Waser, M. Bajdich, S. Nemšák, J.T. Mefford, and W.C. Chueh, "Tuning electrochemically driven surface transformation in atomically flat $LaNiO_3$ thin films for enhanced water electrolysis," Nat Mater **20**(5), 674–682 (2021).

[33] J. Hemberger, A. Krimmel, T. Kurz, H.A. Krug von Nidda, V.Y. Ivanov, A.A. Mukhin, A.M. Balbashov, and A. Loidl, "Structural, magnetic, and electrical properties of single-crystalline (formula presented) (formula presented)," Phys Rev B **66**(9), 1–8 (2002).

[34] J. Gracia, J. Munarriz, V. Polo, R. Sharpe, Y. Jiao, J.W.H. Niemantsverdriet, and T. Lim, "Analysis of the Magnetic Entropy in Oxygen Reduction Reactions Catalysed by Manganite Perovskites," ChemCatChem **9**(17), 3358–3363 (2017).

[35] F. Tsui, M.C. Smoak, T.K. Nath, and C.B. Eom, "Strain-dependent magnetic phase diagram of epitaxial $La_{0.67}Sr_{0.33}MnO_3$ thin films," Appl Phys Lett **76**(17), 2421–2423 (2000).

[36] E. Dagotto, T. Hotta, and A. Moreo, "Colossal magnetoresistant materials: the key role of phase separation," Phys Rep **344**, 1–153 (2001).

[37] L. Yin, C. Wang, Q. Shen, and L. Zhang, "Strain-induced Curie temperature variation in $La_{0.9}Sr_{0.1}MnO_3$ thin films," RSC Adv **6**(98), 96093–96102 (2016).

[38] G. Li, H.D. Zhou, S.J. Feng, X.J. Fan, X.G. Li, and Z.D. Wang, "Competition between ferromagnetic metallic and paramagnetic insulating phases in manganites," J Appl Phys **92**(3), 1406–1410 (2002).

[39] S. Fletcher, "Tafel slopes from first principles," Journal of Solid State Electrochemistry **13**(4), 537–549 (2009).

[40] S. Czioska, K. Ehelebe, J. Geppert, D. Escalera-López, A. Boubnov, E. Saraçi, B. Mayerhöfer, U. Krewer, S. Cherevko, and J.D. Grunwaldt, "Heating up the OER: Investigation of IrO2 OER Catalysts as Function of Potential and Temperature**," ChemElectroChem **9**(19), (2022).

[41] L. Chen, Z. Wang, G. Wang, H. Guo, M. Saghayezhian, Z. Liao, Y. Zhu, E.W. Plummer, and J. Zhang, "Surface and interface properties of $La_{1/3}Sr_{2/3}MnO_3$ thin films on $SrTiO_3$ (001)," Phys Rev Mater **3**(4), (2019).

[42] G. Shibata, K. Yoshimatsu, E. Sakai, V.R. Singh, V.K. Verma, K. Ishigami, T. Harano, T. Kadono, Y. Takeda, T. Okane, Y. Saitoh, H. Yamagami, A. Sawa, H. Kumigashira, M. Oshima, T. Koide, and A. Fujimori, "Thickness-dependent ferromagnetic metal to paramagnetic insulator transition in $La_{0.6}Sr_{0.4}MnO_3$ thin films studied by x-ray magnetic circular dichroism," Phys Rev B **89**(23), (2014).

[43] L. Li, Y. Ji, Z. Diao, J. Zhang, and Z. Liao, "Toward ultrathin ferromagnetic metal of (110) $La_{2/3}Sr_{1/3}MnO_3$ thin films," Appl Phys Lett **117**(12), (2020).

[44] Z. Liao, N. Gauquelin, R.J. Green, S. Macke, J. Gonnissen, S. Thomas, Z. Zhong, L. Li, L. Si, S. Van Aert, P. Hansmann, K. Held, J. Xia, J. Verbeeck, G. Van Tendeloo, G.A. Sawatzky, G. Koster, M. Huijben, and G. Rijnders, "Thickness Dependent Properties in Oxide Heterostructures Driven by Structurally Induced Metal–Oxygen Hybridization Variations," Adv Funct Mater **27**(17), (2017).

[45] S. Macke, J.E. Hamann-Borrero, R.J. Green, B. Keimer, G.A. Sawatzky, and M.W. Haverkort, "Dynamical Effects in Resonant X-Ray Diffraction," Phys Rev Lett **117**(11), (2016).

[46] P. Reith, X. Renshaw Wang, and H. Hilgenkamp, "Analysing magnetism using scanning SQUID microscopy," Review of Scientific Instruments **88**(12), (2017).

[47] J.R. Kirtley, M.B. Ketchen, K.G. Stawiasz, J.Z. Sun, W.J. Gallagher, S.H. Blanton, and S.J. Wind, "High-resolution scanning SQUID microscope," Appl Phys Lett, 1138 (1995).

[48] S. Anantharaj, and S. Noda, "iR drop correction in electrocatalysis: everything one needs to know!," J Mater Chem A Mater **10**(17), 9348–9354 (2022).

[49] D.G. Hawthorn, F. He, L. Venema, H. Davis, A.J. Achkar, J. Zhang, R. Sutarto, H. Wadati, A. Radi, T. Wilson, G. Wright, K.M. Shen, J. Geck, H. Zhang, V. Novák, and G.A. Sawatzky, "An in-vacuum diffractometer for resonant elastic soft x-ray scattering," Review of Scientific Instruments **82**(7), (2011).

[50] L.M.A. Monzon, and J.M.D. Coey, "Magnetic fields in electrochemistry: The Lorentz force. A mini-review," Electrochem Commun **42**, 38–41 (2014).

[51] H. bo Zheng, Y. li Wang, P. Zhang, F. Ma, P. zhao Gao, W. ming Guo, H. Qin, X. pan Liu, and H. ning Xiao, "Multiple effects driven by AC magnetic field for enhanced electrocatalytic oxygen evolution in alkaline electrolyte," Chemical Engineering Journal **426**, (2021).

[52] S. Luo, K. Elouarzaki, and Z.J. Xu, "Electrochemistry in Magnetic Fields," Angewandte Chemie - International Edition **61**(27), (2022).

# Supplementary information

Supplementary discussion A: Further analysis of CV measurements

As mentioned in the main text, cyclic voltammetry measurements were performed, both in the absence and presence of an applied external magnetic field in different in-plane and out-of-plane directions. Selected regions of these measurements for the 13 uc ferromagnetic film are shown Supplementary figure 5. In these graphs, an OER enhancement is visible for the 2 different in-plane directions. To compare the effect of the different field directions on the PM and FM films, the current density at an *iR*-corrected potential of 1.8 V vs RHE are plotted in Figure 5. Here, a systematic enhancement of ~0.4% upon application of an in-plane magnetic field (Figure 5 (a)&(b)) can be observed for the ferromagnetic film when comparing the activity with (colored bars) and without (grey bars) an external magnetic field. For the 10 uc thick paramagnetic film (Figure 5 (e)&(f)), there is no systematic difference with or without applied field along all in-plane directions. Moreover, some ageing and scattering effects were observed for each measurement as changes in the activity are present even for the sweeps where no field was applied (grey bars).

Supplementary discussion B: Magnetic field induced OER enhancement in a paramagnetic thin film

As shown in the main text, the paramagnetic sample showed an unexpected behavior in contradiction with previously published results [6,12,21] as distinguishable changes in current density upon magnetic field exposure were observed. The fact that these effects have been previously overlooked could be explained by the lack of investigation of directional dependence or the small size of the effects (~0,1 %). A possible origin of the current changes can be found in Lorentz forces affecting the charged species in the catalyst which are moving in the magnetic field due to the rotation of the working electrode (see Supplementary figure 7). [21,50,51] The magneto-hydrodynamic effect can be excluded as the current density enhancements showed no systematic dependence on the stirring behavior (Supplementary figure 8). [51,52]

Supplementary discussion C: Possible contributions to the temperature-dependent OER activity in ferromagnetic films

To solidify the claim that a change of intrinsic magnetic order at the transition temperature in ferromagnetic is the main origin for the OER enhancement, possible competing origins have to be excluded.

Sample degradation and surface restructuring were ruled out as an origin of the OER activity as measurements done using atomic force microscopy and X-ray reflectivity before and after OER showed only minor and similar changes for both films (Supplementary figure 9 and Supplementary figure 10). To pin-point which other factors could be of importance we combine this observation with the hypothesis that purely temperature-dependent electrochemical effects should be similar for both films. From this we conclude that only the (magneto-) resistance and magnetic order and the temperature dependence of these effects should be significantly different for both samples investigated. Only these factors can thus induce the activity enhancement, which counteracts the temperature-induced activity decrease.

To investigate whether the difference in (magneto-)resistance in the catalytic films could be the origin of the enhancement a closer look has to be taken at the data shown in Figure 7(c) and Supplementary figure 1(b),(c) and (d). It can be observed that the resistance ratio of $R^{\text{FM sample}}$ to $R^{\text{PM sample}}$ does not abruptly change at $T_c$ whereas the difference in magnetization between the two films changes abruptly above $T_c$ and keeps increasing with decreasing temperature, similar to the continuous enhancement of the current density of the ferromagnetic sample compared to the paramagnetic sample (Figure 7 (a)&(b) and Supplementary figure 1(b)). Furthermore, the relative increase of $R^{\text{FM sample}}$ with increasing temperature is only ~20% larger than that of $R^{\text{PM sample}}$ from 295-335K (Supplementary figure 1(d)) compared to the 35% enhancement in OER current. A lowering of the resistance in the ferromagnetic sample due to spin interactions can thus not be the sole reason of an enhancement of the current density.

Supplementary Table: Reported OER enhancements in literature

| Ref | Material | Maximal reported enhancement* | Origin for increment difference |
|---|---|---|---|
| 6 | $CoFe_2O_4$ | ~ 100% at 1.7 V vs RHE | Larger fields utilized, higher coercivity and higher saturation magnetization |
| 11 | Ni foil | ~ 50% at 1.7 V vs RHE | Larger fields utilized and seems that some irreversible changes in the catalyst induce enhancement |
| 12 | Ni foil | ~ 50% at 1.7 V vs RHE | Larger fields utilized and very different material structure |
| 13 | $NiZnFe_4O_x$ | ~ 100% at 1.65 V vs RHE | Larger fields utilized and very different material structure |
| 14 | $CoFe_2O_4$ | ~ 100% at 1.65 V vs RHE | Larger fields utilized and higher coercivity (For small coercive films in this paper enhancement lot smaller) |
| 16 | Ni single atom in $MoS_2$ host | ~ 500% at 1.8 V vs RHE | Larger fields utilized and very different material structure |
| 17 | $CoFe_2O_4$/ $Co(Fe)O_xH_y$ | ~ 30% at 1.55V vs RHE | Larger fields utilized, higher coercivity |
| 18 | $Fe_3O_4$@Ni(OH) core shell particles | 43.8% at 1.6V vs RHE | Larger fields utilized |
| 19 | $La_{0.67}Sr_{0.33}MnO_3$ | ~ 100% at 2.0 V vs RHE | Larger fields utilized |
| 20 | $La_{0.7}Sr_{0.2}Ca_{0.2}MnO_3$ | ~ 80% at 1.8 V vs RHE | Larger fields utilized |
| 23 | $Co_xNi_{1-x}Fe_2O_4$-nanosheets | ~ 10% at 1.7 V vs RHE | Larger fields utilized and higher saturation magnetization, moreover smaller enhancement |
| 24 | LiCoVO4 compared to ZnCoVO4 | > 100% at 1.6 V vs RHE | Also change in composition, not clear what the extend of magnetic induced effect is. |
| 25 | $Fe_{0.5}Co_{0.5}$-MOF | > 100% upon at 1.6 V vs RHE | Very different system and approach to induce long-range magnetic coupling |
| 26 | NiFe-LDHs | ~ 10% at 1.6 V vs RHE | Larger fields utilized |
| 27 | $LaCoO_3$ | > 100% at 1.8 V vs RHE | Also effect of surface termination can play a role |

*Supplementary table 1 Estimated reported spin polarized OER enhancements in literature*

*approximated from reported figures

Supplementary figures

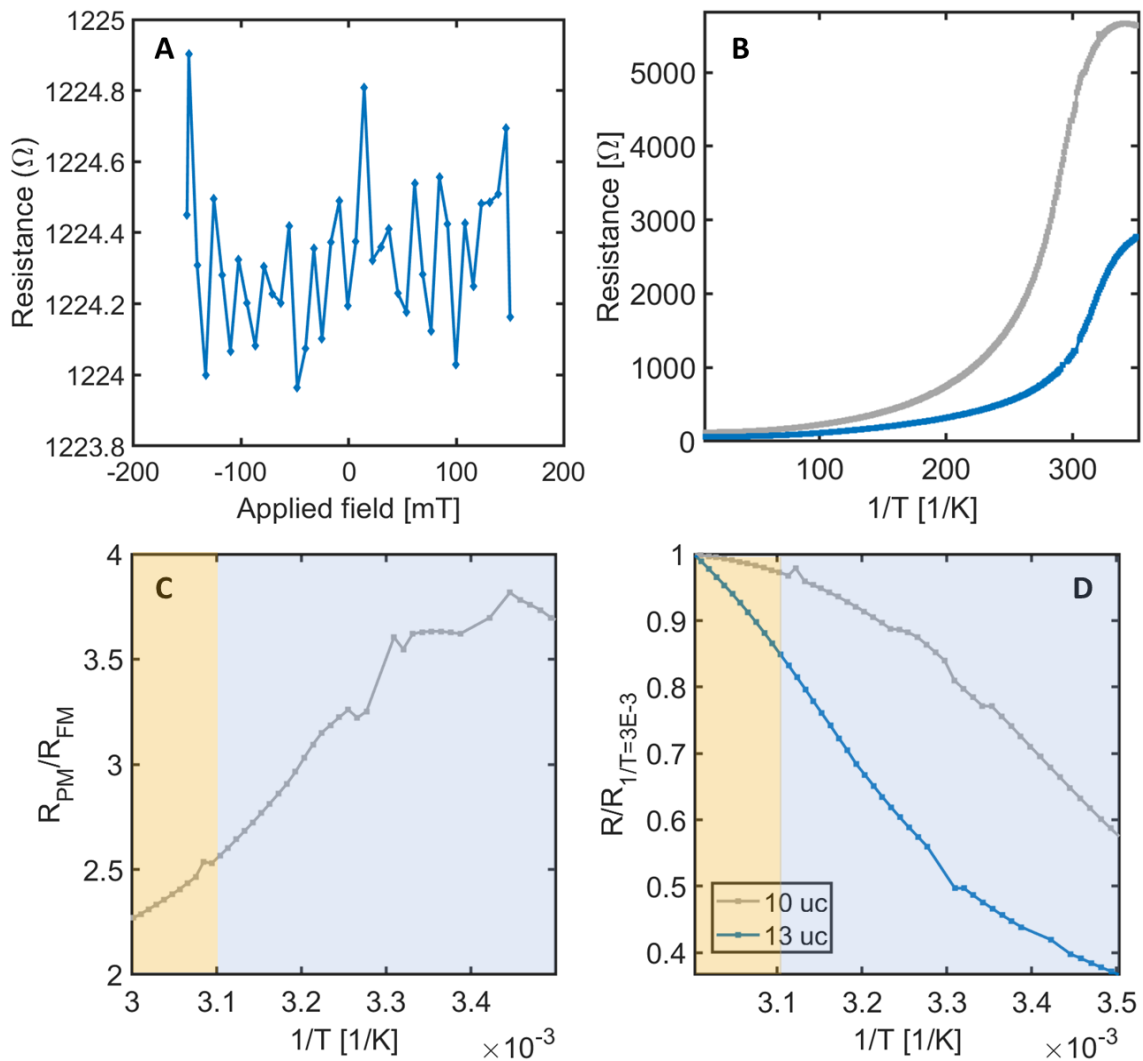

*Supplementary figure 1. Resistance behavior of $La_{0.7}Sr_{0.3}MnO_3$ thin films. (a) Resistance versus applied field for a 13 uc thick film. (b) Resistance versus temperature for a 10 and a 13 uc thick film. (c) Ratio between the resistance of the paramagnetic (10 uc) and ferromagnetic (13uc) thin films versus 1/T. (d) Relative resistance compared to the resistance at $1/T=3\times10^{-3}$ 1/K for a 10 and 13 uc thick film.*

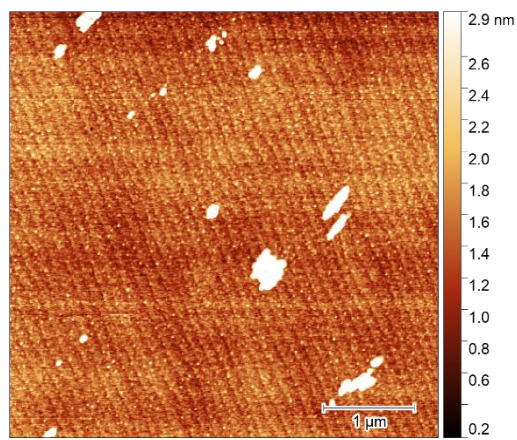

*Supplementary figure 2 Topography image of MFM scanned region of the 13 uc LSMO film*

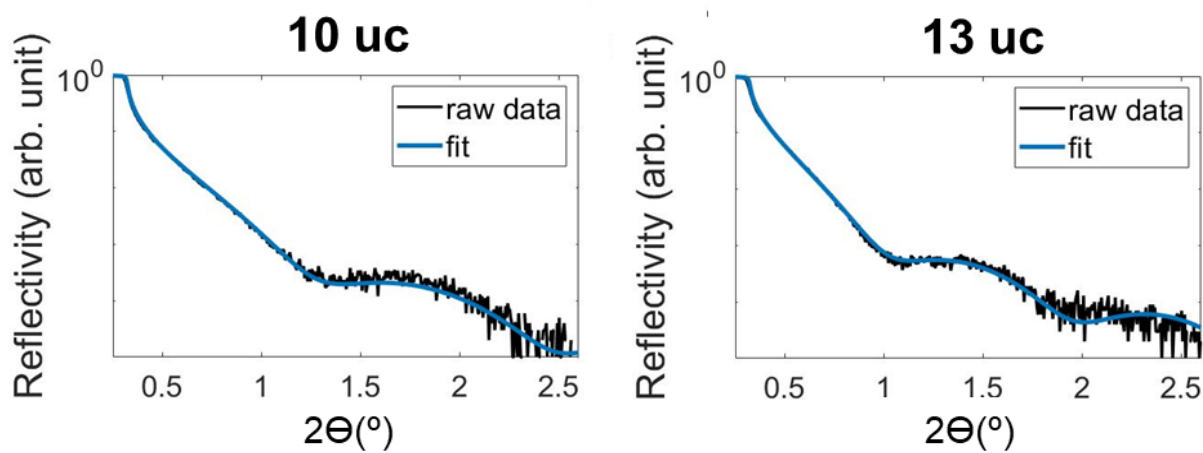

*Supplementary figure 3. X-ray reflectivity measurements of 10 and 13 uc thick films along with a fit of the data. The fits give a thickness of respectively 3.6 and 4.8 nm which is equal to approximately 10 and 13 uc of $La_{0.67}Sr_{0.33}MnO_3$.*

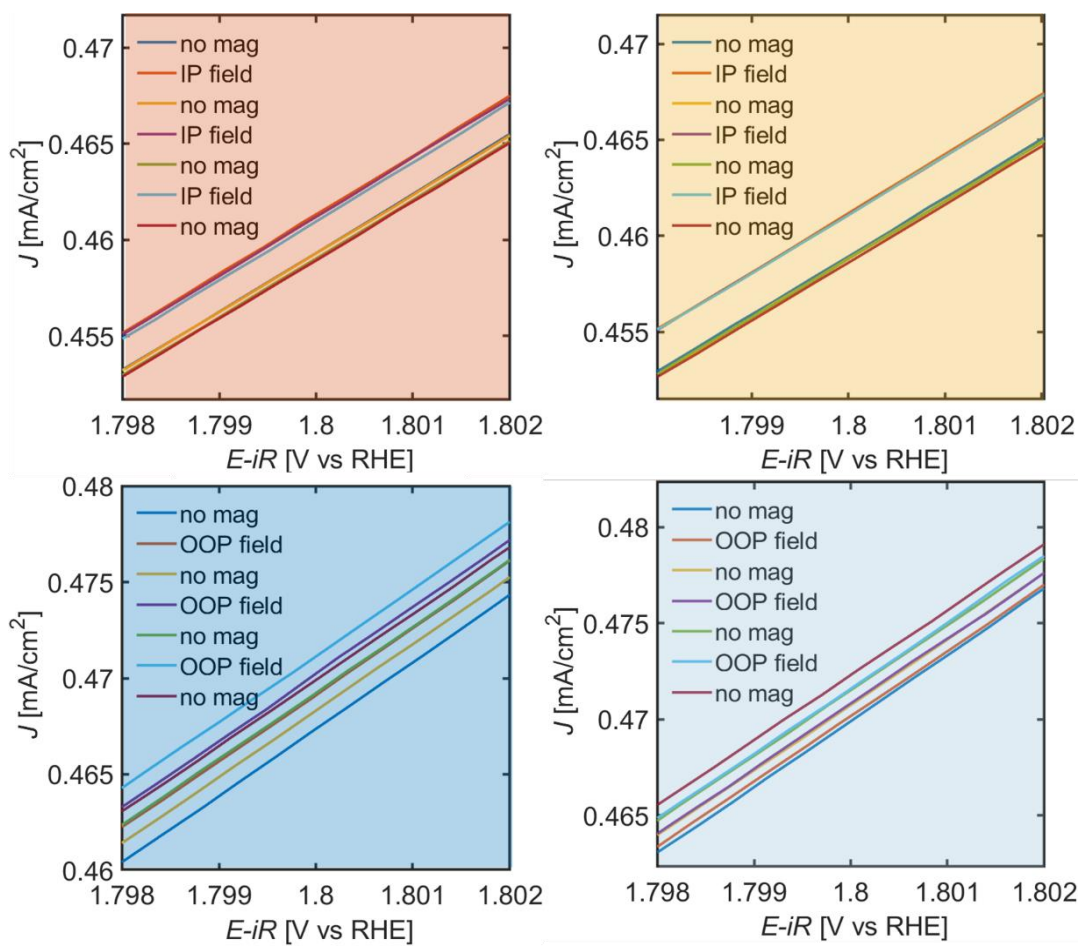

*Supplementary figure 4. Cyclic voltammetry sweeps used to generate the bar graphs shown in Figure 5 for the 13 uc thick $La_{0.67}Sr_{0.33}MnO_3$ film. The colors indicate the direction of the field lines as shown in Figure 4*

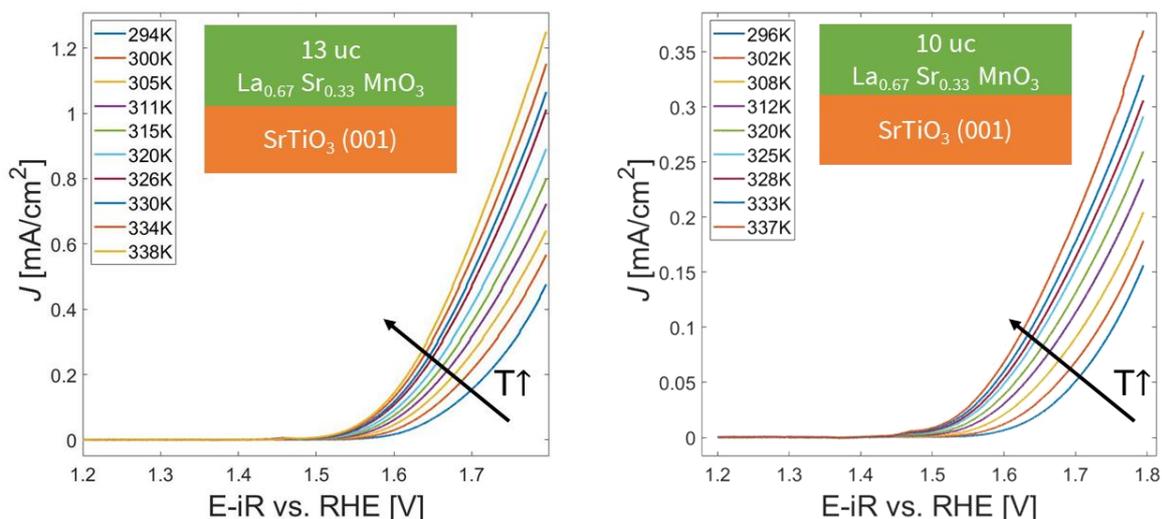

*Supplementary figure 5. Average of the backward and forward sweep of 2nd cycle of the cyclic voltammetry measurement performed at each temperature during T dependent measurements. It should be noted that the scale of the Y-axis is different for both samples.*

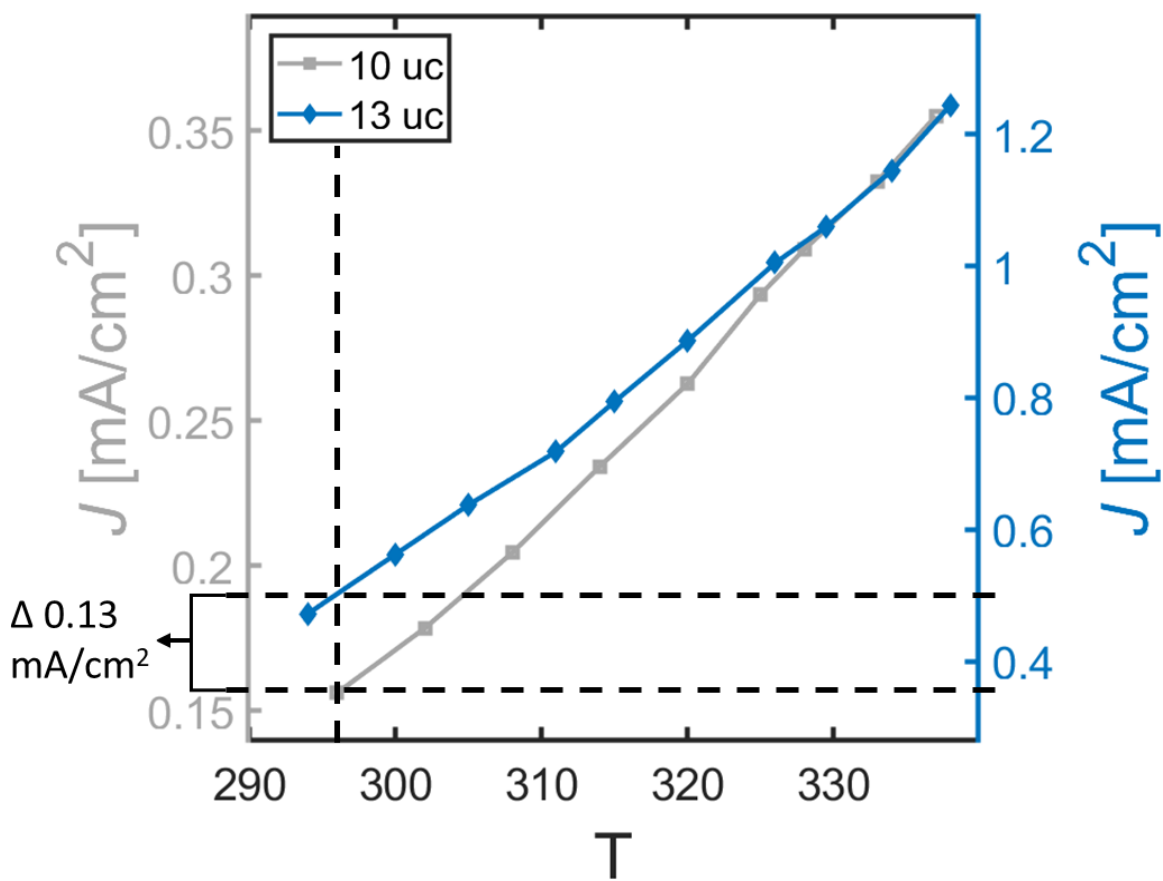

*Supplementary figure 6. Plot of the current density versus temperature obtained for a ferromagnetic (13 uc) and a paramagnetic (10 uc) $La_{0.67}Sr_{0.33}MnO_3$ film at an iR-corrected potential of 1.8 V vs RHE. To obtain an upper estimate of the current density, we assumed that the behavior of the 10 uc thick film scaled to the current density of the 13 uc film could predict the purely temperature dependent behavior of the 13 uc thick film. We then computed the difference between this curve and the measured curve of the 13 uc thick film. This difference of 0.13 mA/cm$^2$ is equal to an increase of 35% for the assumed non magnetic curve.*

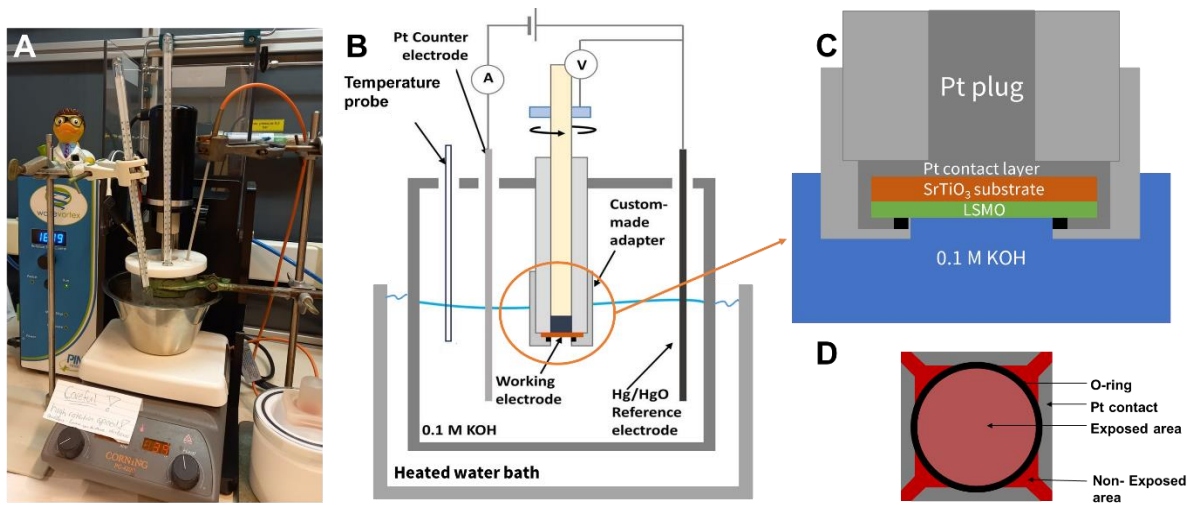

*Supplementary figure 7. (a) picture of the used electrochemical setup. (b) schematic representation of the setup to the left. (c) schematic representation of the circled part in B (not to scale). (d) Sputtered front side of a thin film sample with the O-ring placed on top. The area exposed to the electrolyte is shaded and equal to 0.441 cm$^2$.*

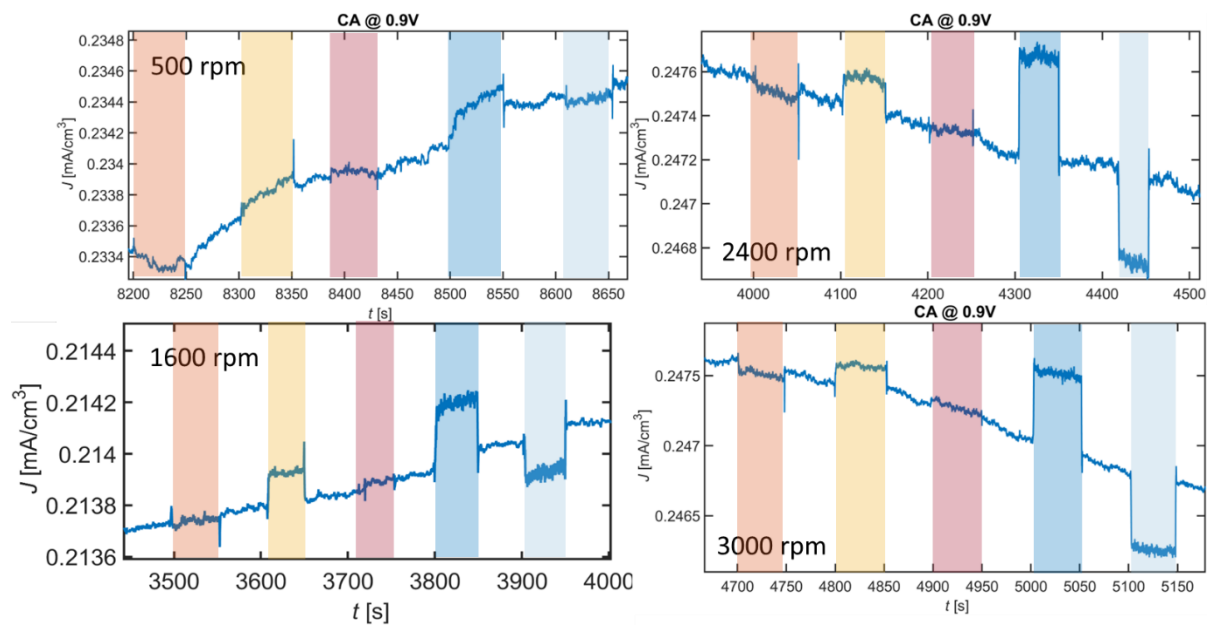

*Supplementary figure 8. Rotation speed dependence of the magnetic enhancement upon external field exposure during chronoamperometry on a 13 uc thick La$_{0.67}$Sr$_{0.33}$MnO$_3$ film. As can be seen there is a large difference between the measurement done at 500 rpm and the rest, this is due to the fact that here the low rotation speed leads to the formation of a lot of bubbles and diffusion limitations. However, for the rest of the measurements a very similar behavior is observed upon field exposure. As we normally operate at 1600 rpm, we can thus conclude that a slight enhancement of diffusion of ionic species in the liquid through the MHD effect is not the origin for the enhancement in current density seen when applying a field.*

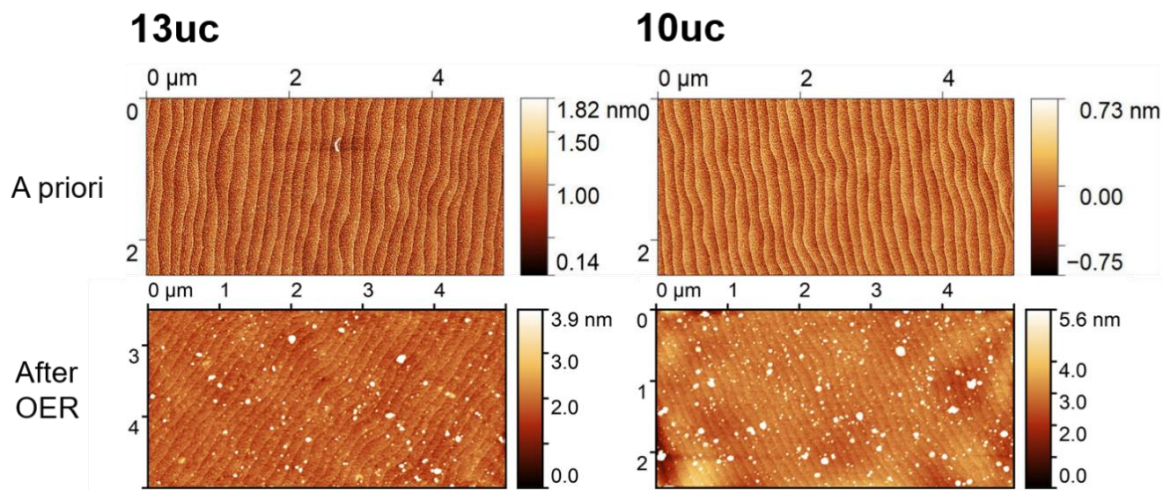

*Supplementary figure 9. Atomic force microscopy of both samples directly after growth (a priori) and after electrochemical measurements. As can be seen, the morphology is very similar except for some small particles which are likely induced by drying of the electrolyte on the sample*

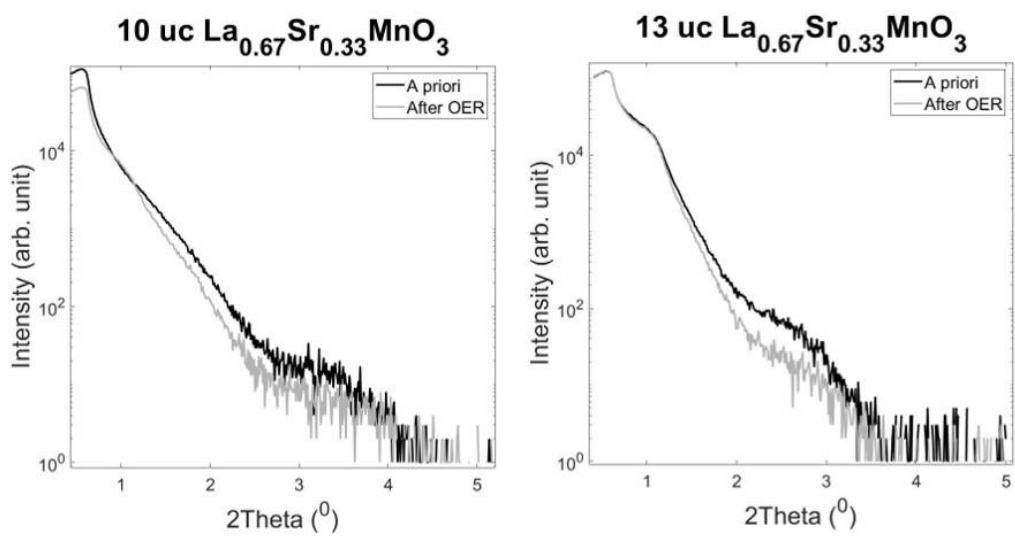

*Supplementary figure 10. X-ray reflectivity of both samples directly after growth (a priori) and after electrochemical measurements. As can be seen, the location of the first oscillation does not change. We conclude that the thickness of the films does not change appreciably. The decrease in intensity is induced by the contamination of dried electrolyte.*

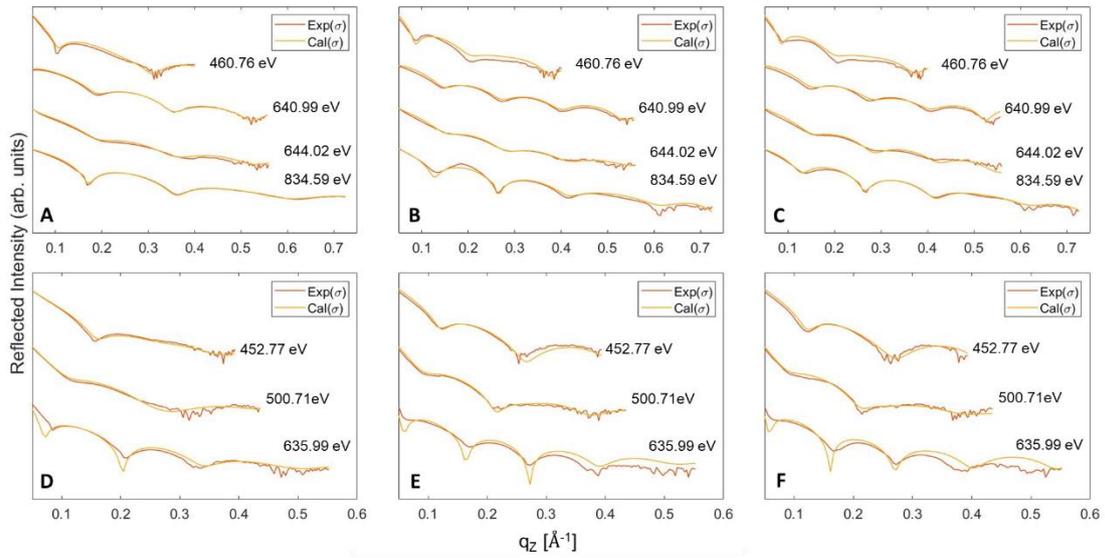

*Supplementary figure 11. Resonant theta/two-theta reflectivity scans of $La_{0.7}Sr_{0.3}MnO_3/SrTiO_3$ for samples of (a) the 10 uc film measured at a temperature of 150 K, (b) the 13 uc film measured at a temperature of 150 K, and (c) the 13 uc film measured at a temperature of 300 K. The non-resonant theta/two-theta reflectivity scans are illustrated for the same heterostructures of (d) the 10 uc film measured at a temperature of 150 K, (e) the 13 uc film measured at a temperature of 150 K, and (f) the 13 uc film measured at a temperature of 300 K. The experimental and simulated data are included in each plot, where each pair is labeled with their corresponding photon energy. The chosen photon energies are indicated in Supplementary figure 13. Each set of curves is offset vertically for clarity.*

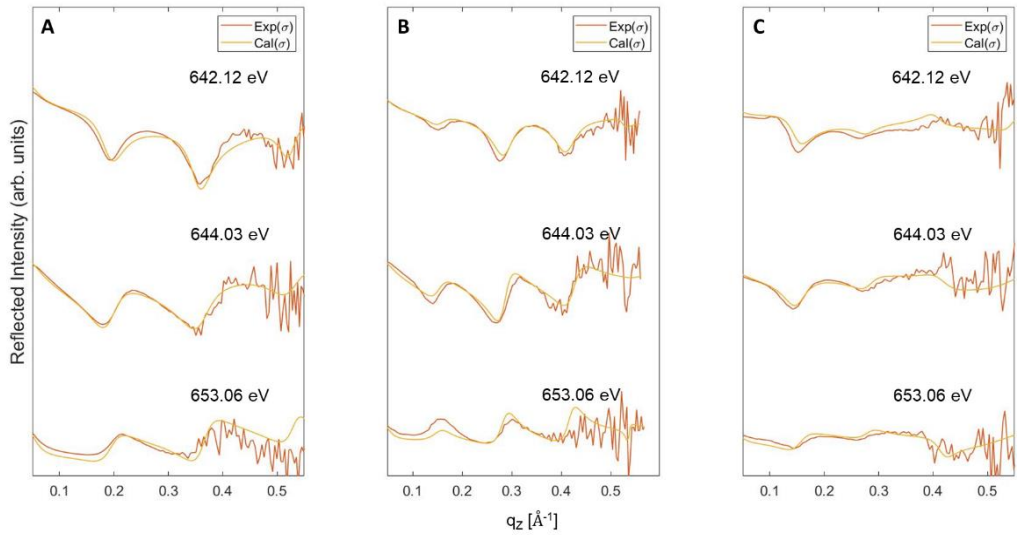

*Supplementary figure 12. circular polarized asymmetry curves of $La_{0.7}Sr_{0.3}MnO_3/SrTiO_3$ for samples of (a) the 10 uc film measured at a temperature of 150 K, (b) the 13 uc film measured at a temperature of 150 K, and (c) 13 uc film measured at a temperature of 300 K. The experimental and simulated data pairs are included in each plot providing a label for each pair's corresponding photon energy. Each set of curves is offset vertically for clarity.*

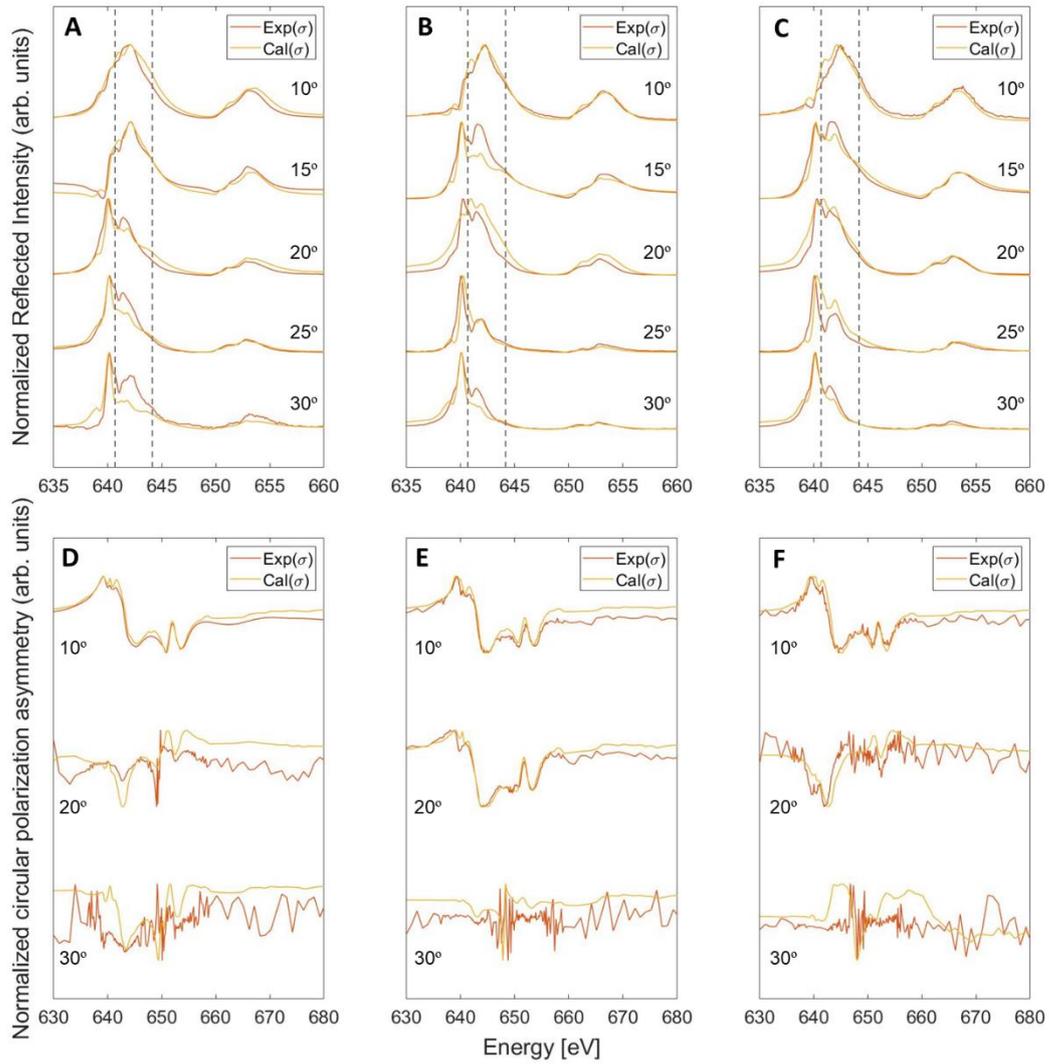

*Supplementary figure 13. Mn-resonant energy scans of La$_{0.7}$Sr$_{0.3}$MnO$_3$/SrTiO$_3$ samples for (a) the 10 uc film measured at a temperature of 150 K, (b) the 13 uc film measured at a temperature of 150 K, and (c) the 13 uc film measured at a temperature of 300 K. Vertical dashed lines indicate the energy for the on-resonance scans in Supplementary figure 11. Mn-resonant circular polarized asymmetry spectra of La$_{0.7}$Sr$_{0.3}$MnO$_3$/SrTiO$_3$ samples for (d) the 10 uc film measured at a temperature of 150 K, (e) the 13 uc film measured at a temperature of 150 K, and (f) the 13 uc film measured at a temperature of 300 K. The experimental and simulated data pairs are included in each plot providing a label for each pair's corresponding grazing angle. The results were normalized and each set of curves is offset vertically for clarity*